\def\bsax{{\it Beppo}SAX\ }
\def\swift{{\it Swift\ }}
\def\swiftns{{\it Swift}}
\def\hete{{\it HETE}\ }
\def\hetetwo{{\it HETE-2}\ }
\def\rxte{{\it RXTE}\ }
\def\eop{E^{\rm obs}_{\rm peak}}
\def\eiso{E_{\rm iso}}

\def\ep{E_{\rm peak}}
\def\egamma{E_\gamma}
\def\se{S_E}

\def\thetajet{\theta_{\rm jet}}
\def\omegam{\Omega_M}
\def\omegal{\Omega_{\rm DE}}
\def\tjet{t_{\rm jet}}

\documentclass[11pt,preprint]{aastex}

\begin{document}
\pagestyle{empty}

\title{{\bf A Gamma-Ray Burst Mission to Investigate the Properties
of Dark Energy \\}
\medskip
{\it A White Paper Submitted to the Dark Energy Task Force}\\
{\it 15 June 2005}\\
\medskip
\small{{\bf Point of contact: d-lamb@uchicago.edu}}}

\noindent
D. Q. Lamb$^1$, G. R. Ricker$^2$, D. Lazzati$^3$, G. Ghirlanda$^4$, G.
Ghisellini$^4$, C. Firmani$^{4,5}$, L. Amati$^6$, J.-L. Atteia$^7$, V.
Avila-Reese$^5$, S. Burles$^2$, N. Butler$^2$, H.-W. Chen$^2$, E.
Costa$^{8}$, J. Doty$^{9}$, F. Frontera$^6$, A. Fruchter$^{10}$, P.
Garnavich$^{11}$, C. Graziani$^1$, J. G. Jernigan$^{12}$, N.
Kawai$^{13}$, P. Mazzali$^{14}$, P. Meszaros$^{15}$, L. Piro$^{8}$, T.
Sakamoto$^{16}$, K. Stanek$^{17}$, M. Vietri$^{18}$, M. della
Valle$^{19}$, J. Villasenor$^2$, B. Zhang$^{20}$

\vspace{0.2in}
\noindent 
$^1$U. Chicago; $^2$MIT; $^3$U. Colorado; $^4$Osservatorio Astronomico
di Brera; $^5$U.N.A.M.; $^6$IAFSC, sezione Bologna, CNR; 
$^7$Observatoire Midi-Pyr\'en\'ees; $^8$Istituto di Astrofisica
Spaziale e Fisica Cosmica -- INAF, Roma; $^9$Noqsi Aerospace, Ltd.;
$^{10}$STScI; $^{11}$U. Notre Dame; $^{12}$UC Berkeley; $^{13}$Tokyo
Institute of Technology; $^{14}$Osservatorio Astronomico, Trieste; 
$^{15}$Penn State; $^{16}$GSFC; $^{17}$CfA; $^{18}$Scuola Normale
Superiore, Pisa; $^{19}$Osservatorio Astronomico di Arcetri;
$^{20}$UNLV

\newpage
\pagestyle{plain}
\pagenumbering{arabic}
\setcounter{page}{1}
\centerline {\bf Executive Summary}
%\vspace{0.2in}

Gamma-ray bursts (GRBs) are the most brilliant events in the universe. 
The intrinsic luminosities of the bursts span more than five decades. 
At first glance, therefore, these events would hardly seem to be
promising standard candles for cosmology.  However, very recently,
relations between the peak energy $\ep$ of the burst spectrum in $\nu
F_\nu$, the isotropic-equivalent energy $\eiso$ of the burst, and the
radiated energy $\egamma$ of the burst -- all in the rest frame of the
burst source -- have been found. In a way that is exactly analogous to
the way in which the relation between the peak luminosity and the
rate of decline of the light curve of Type Ia supernovae 
can be used to make Type Ia supernovae excellent standard candles
for cosmology, so too, the relations between $\ep$, $\eiso$, and
$\egamma$ point toward a methodology for using GRBs as excellent
standard candles for cosmology.  In addition, GRBs occur over the broad
redshift range from $z=0.1$ to at least $z=4.5$, and both they and their
afterglows are easily detectable out to $z > 8$.  Thus GRBs show great
promise as cosmological ``yardsticks'' to measure the rate of expansion
of the universe over time, and therefore the properties of dark energy
(i.e., $\omegam$, $\omegal$, $w_0$, and $w_a$).

Measurements of four observational quantities are needed in order to
use a GRB as a standard candle'':  (1) the observed peak energy
$\eop$ of the burst spectrum in $\nu F_\nu$ (2) the energy fluence
$\se$ of the burst, (3) the jet break time $\tjet$ of the burst
afterglow, and (4) the redshift $z$ of the burst.  We describe a
concept for a possible MIDEX-class mission dedicated to using GRBs to
constrain the properties of dark energy that would obtain these
quantities for $>$ 800 bursts in the redshift range $0.1 \lesssim z
\lesssim 10$ during a 2-year mission.  This burst sample would enable
both $\omegam$ and $w_0$ to be determined to $\pm 0.07$ and $\pm 0.06$
(68\% CL), respectively, and $w_a$ to be significantly constrained. 
Probing the properties of dark energy using GRBs is complementary (in
the sense of parameter degeneracies) to other probes, such as CMB
anisotropies and X-ray clusters.  GRBs are complementary to Type Ia
supernovae because of the broader redshift range over which GRBs occur,
and the ease with which they and their afterglows can be observed.

The MIDEX-class mission concept calls for three small spacecraft placed
at L$_2$ in the Earth-Sun system.  One spacecraft, a minisatellite, would
carry an array of X-ray CCD cameras very similar to the Soft X-Ray
Camera (SXC) on \hetetwo and a set of large-area NaI detectors.  The
array of X-ray CCD cameras would have a sensitivity $\sim $ twice that
of the Wide-Field X-Ray Monitor (WXM) on \hetetwo. The field of view
(FOV) of the array would be $\approx 2 \pi$ steradians, 
10 $\times$ greater than that of the WXM.  It would
localize bursts to $\approx 1$\arcsec accuracy.  The set of NaI
detectors, combined with the array of X-ray CCD cameras,  would
determine $\eop$ and $\se$ to 10\% accuracy (90\% CL) over the 1--1000
keV energy range.  The other two spacecraft--both microsatellites--would
carry identical, high throughput, short (80 cm) focal length X-ray
telescopes for monitoring the X-ray afterglows of each of the bursts
and determining $\tjet$ for them.  The mission concept incorporates
six dedicated, ground-based 2-meter class telescopes with integral field
spectrographs (IFS) distributed in longitude around the Earth for
promptly observing the optical afterglows of the bursts and determining
their redshifts.

\vfill
\eject
\twocolumn

\vspace{-0.1in}
\section{Precursor Observations}
 
\begin{figure*}
\centerline{
\includegraphics[scale=0.95,clip=]{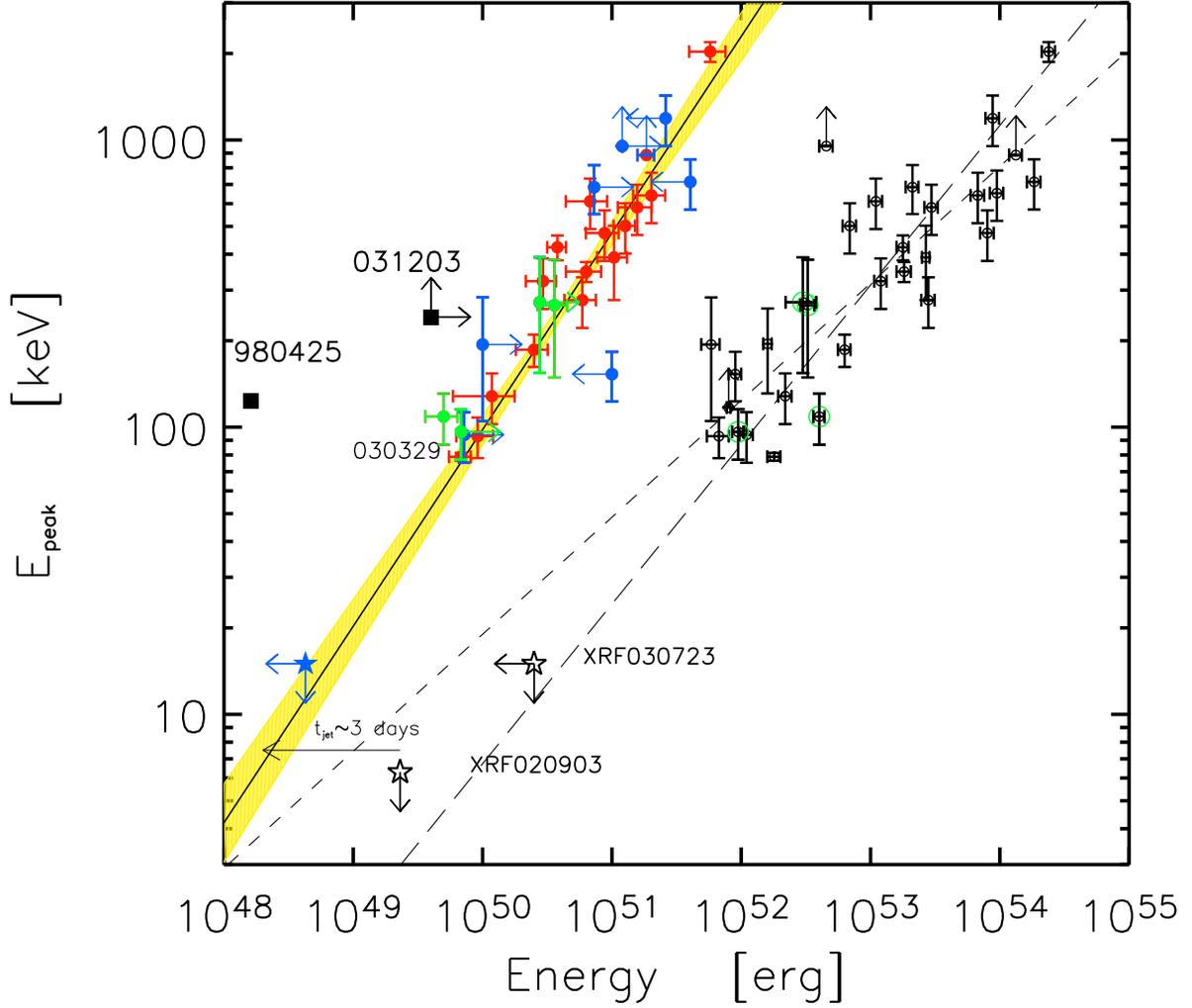}}
\caption{The rest frame $E_{\rm  peak}$--$E_{\rm iso}/E_\gamma$ plane. 
Black open symbols represent the isotropic equivalent energy.  Red
filled symbols are the 15 GRBs for which a jet break was measured in
their afterglow light curves.  Blue
symbols are upper/lower limits for $E_\gamma$. The four new GRBs are
represented as open green circles for $E_{\rm iso}$ and filled green
symbols for $E_{\gamma}$. Also  shown are two outliers (black squares)
for either the Amati and the Ghirlanda correlation (filled squares). 
Stars are the two XRF with known redshift.  The Amati correlation is also
reported either fitting with the errors on both coordinates (long
dashed line) and with the least square method (dashed line). The best
fit Ghirlanda correlation (solid black line), giving a reduced
$\chi^2=1.33$ and a slope $\sim0.7$, is also shown with its
uncertainty region (shaded area).  (Figure from \citep{gglf2005}).}
\label{fig:ghirlanda_3_fig1}
\end{figure*}

Recently, \cite{amati2002} discovered a relation between the peak
energy $\ep$ of the burst spectrum in $\nu F_\nu$ and the
isotropic-equivalent energy $\eiso$ of the burst in the source rest
frame, using 10 \bsax GRBs with known redshifts.  Lamb et al. (2004;
see also Sakamoto et al. 2004, 2005) confirmed this relation for GRBs
and found evidence that it extends down to X-ray-rich GRBs (XRRs) and
X-Ray Flashes (XRFs) and therefore spans five decades in $\eiso$, 

Very recently, \cite{ggl2004} provided further confirmation of this
relation and discovered that an even tighter relation exists between
$\ep$ and $\egamma$, where $\egamma \equiv (1-\cos \thetajet) \eiso$
(see Figure \ref{fig:ghirlanda_3_fig1}), where $\thetajet$ is inferred
from the jet break time $\tjet$.  Still more recently, \cite{liang2005}
have shown that an equally tight relation exists directly among $\ep$,
$\eiso$, and $\tjet$, which places the $\ep-\egamma$ relation on an
entirely empirical footing.   

Using the $\ep-\egamma$ relation, \cite{gglf2004} and \cite{fgga-v2005}
have shown that GRBs can be used as standard candles for cosmology.  
In, particular, they have shown that the uncertainty in $\eiso$ using
this method is currently about 0.25 dex, which is only twice the
current uncertainty in the luminosities $L_{SN}$ of Type Ia supernovae
(SNe) after using the analogous relation between $L_{SN}$ and the rate
of decline of the SN light curves \cite{riess2004}.  These results
suggest that, although the use of GRBs as standard candles is less than
a year old, while the use of Type Ia SNe is more than a half century
old, GRBs hold great promise for cosmology.  GRBs are also
complementary to Type Ia SNe because of the broad redshift range over
which GRBs occur, and the ease with which they and their afterglows can
be observed.

\begin{figure*}
\centerline{
\includegraphics[scale=0.75,clip=]{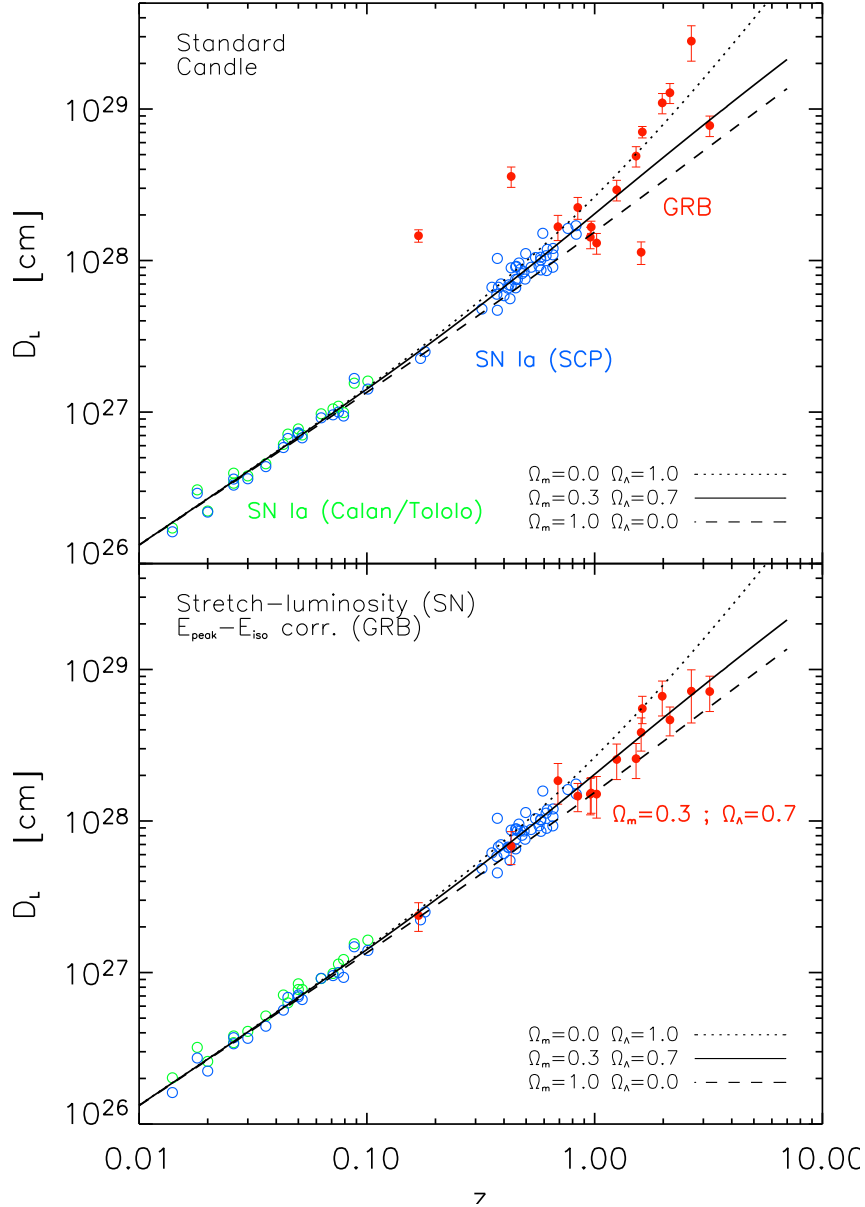}}
\caption{Classical Hubble diagram in the form of luminosity--distance
$D_{\rm L}$ vs. redshift $z$ for Supernova Ia (open green circles:
Cal\`an/Tololo  sample (Hamuy et al. 1996); open blue circles: 
Perlmutter et al. 1999) and GRBs (filled red circles: the 15 bursts in
Ghirlanda et al. 2004b).   In the top panel the SN Ia and GRBs  are
treated as standard candles (no corrections applied); for GRBs
$E_\gamma=10^{51}$ erg is assumed.  In the bottom  panel, we  have 
applied the stretching--luminosity and the $E_\gamma$--$E_{\rm peak}$
relations to SN Ia and GRBs, respectively, as explained in the text. 
Note that, for GRBs, the applied correction depends upon the assumed
cosmology: here we assume the standard $\omegam=0.3$,
$\omegal=0.7$~cosmology.  Both panels also show different $D_{\rm
L}(z)$ curves, as labelled. 
>From \cite{gglf2004}.}
\label{fig:ghirlanda_2_fig1}
\end{figure*}

\begin{figure*}
\centerline{
\includegraphics[scale=0.95,clip=]{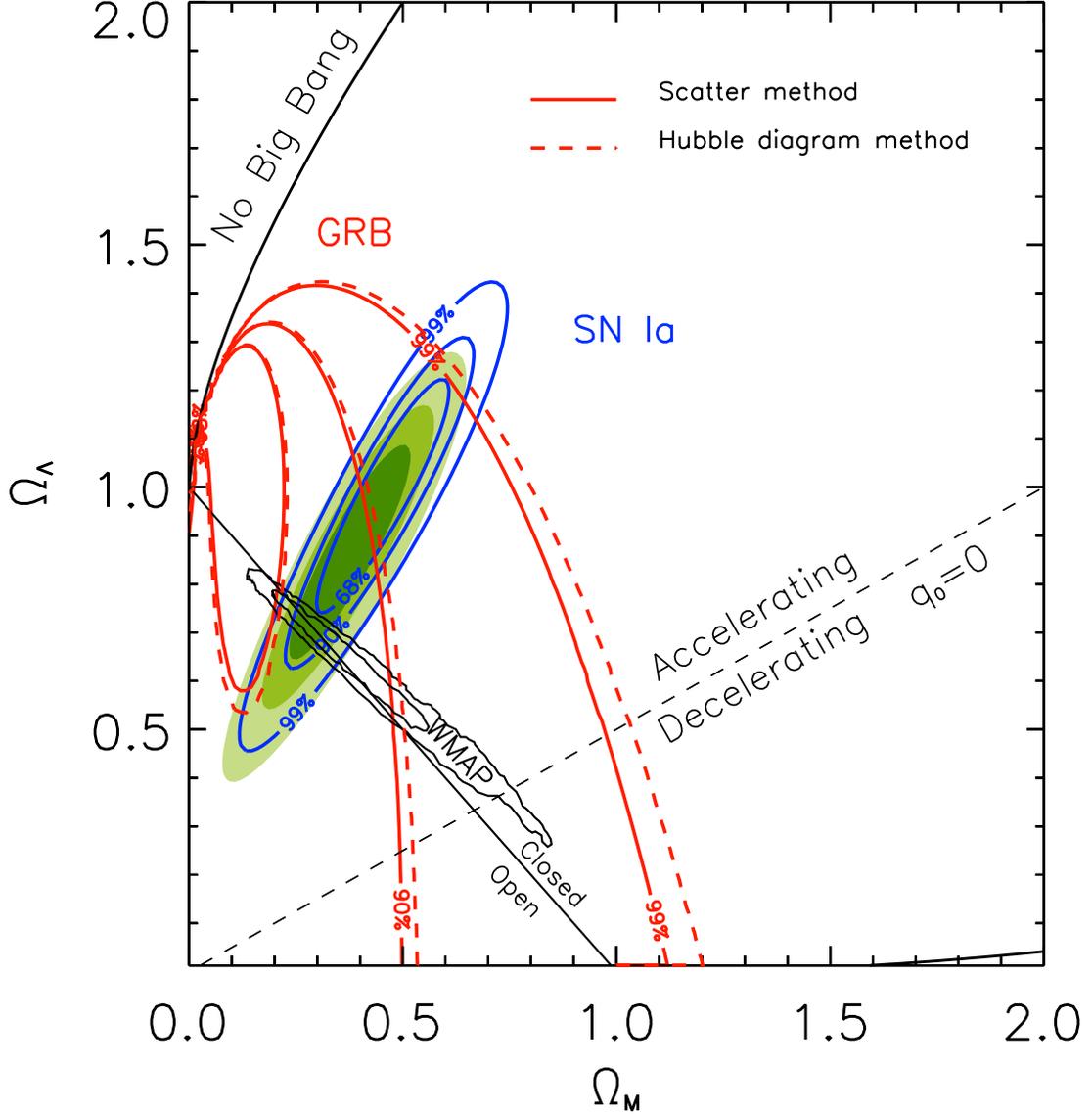}}
\caption{Constraints in the $\omegam-\omegal$~($\omegal=\Omega_{\Lambda}$)
plane derived for our GRB
sample (15 objects, red contours);  the ``Gold'' Supernova Ia sample of
R04) (156 objects, blue contours, derived assuming a
fixed value of $H_0=65$ km s$^{-1}$ Mpc$^{-1}$,  making the contours
slightly different from Fig. 8 of R04). The WMAP
satellite constraints (black contours, Spergel et al. 2003) are also
shown. The three colored ellipsoids are the confidence regions (dark
green: 68\%; green: 90\%; light green: 99\%)  for the combined fit of
SN Ia and our GRB sample. For GRBs only, the minimum
$\chi_{red}^2=1.04$, is at $\omegam=0.07$, $\omegal=1.2$.  From
\cite{gglf2004}.}
\label{fig:ghirlanda_2_fig2}
\end{figure*}

\begin{figure*}
\centerline{
\includegraphics[scale=0.95,clip=]{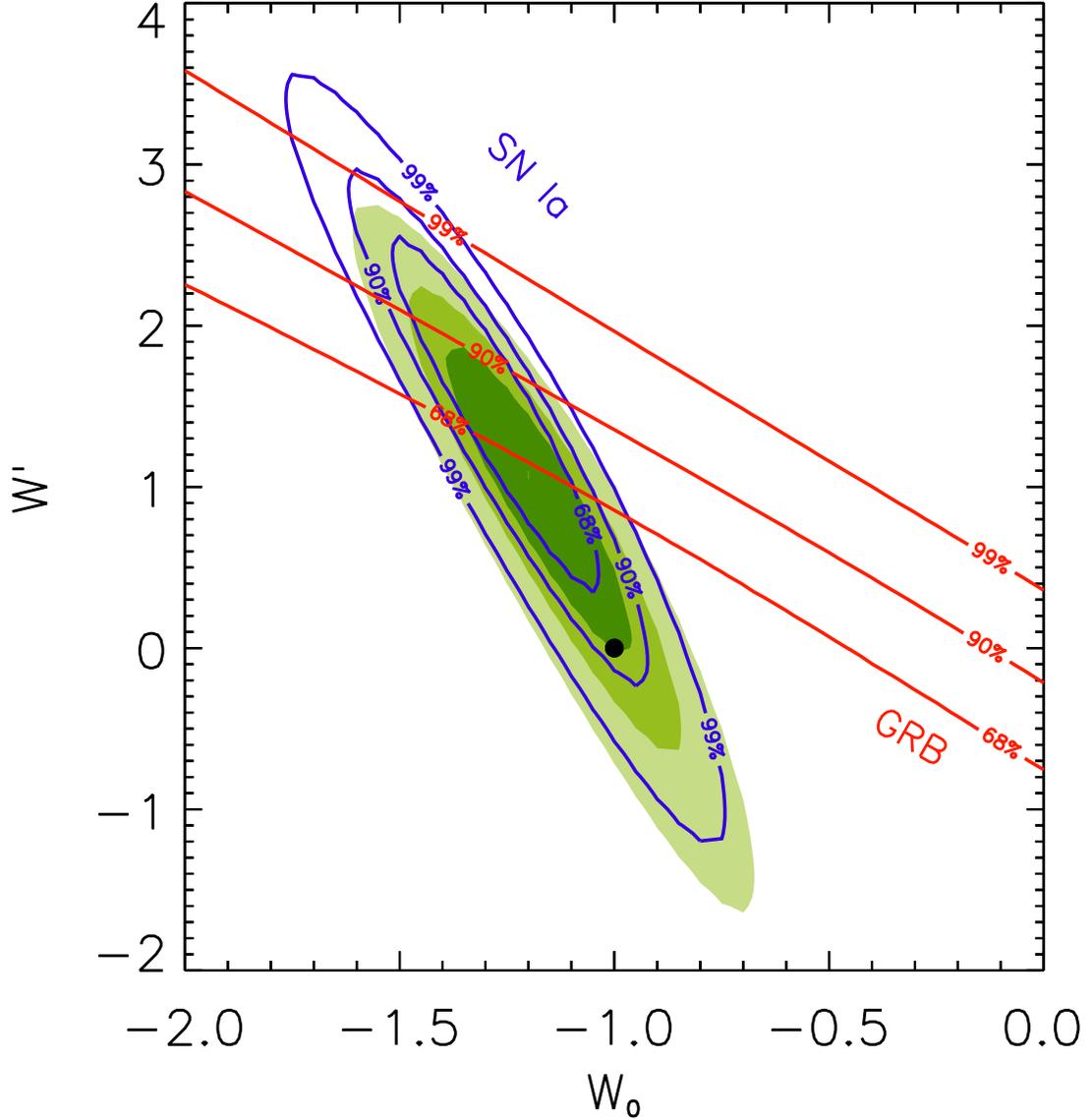}}
\caption{Constraints on the $w_0$, $w^\prime$ parameters entering the
equation of state $p=(w_0+w^\prime z)\rho c^2$, where $\rho$ is the
dark energy density. $w_0=-1$ and $w^\prime=0$ correspond to the
cosmological constant $\omegal$. We assume a flat geometry and
$\omegam=0.27$ (see
also R04). Blue contours: constraints from type Ia SN (R04). Red
contours: constraints from our GRBs, Colored regions: combined
constraints (dark green, green and light green for the 68\%, 90\% and
99\% confidence levels, respectively).  From \cite{gglf2004}.}
\label{fig:ghirlanda_2_fig3}
\end{figure*}

The ``Gold'' GRB sample used by Ghirlanda et al. (2004; see also
Ghisellini et al. 2005 and Firmani et al. 2005) consists of 15 GRBs
for which $\ep$, $\eiso$, $\tjet$, and $z$ are well determined. 
\cite{gglf2004} and \cite{fgga-v2005} show that, using this small
sample of GRBs, interesting constraints can be placed on $\omegam$,
$\omegal$, $w_o$, and $w_1$, assuming $w(z) = w_0 + w_1 z$ (see
Figures \ref{fig:ghirlanda_2_fig1}-\ref{fig:ghirlanda_2_fig3}).  

Precursor observations are needed to determine $\eop$ $\se$, $\tjet$,
and $z$ for a much larger sample of GRBs, and to use this much larger
sample of bursts to confirm that (1) most GRBs satisfy the
$\ep-\eiso-\tjet$ relation, (2) this relation extends down in
$\eiso$ and $\ep$ to XRFs, (3) the systematic errors in this relation
are small, and (4) X-ray afterglow observations can accurately
determine $\tjet$ for most bursts, including XRFs, for which $\tjet$
can be as large as $\sim$ 20 days.

Confirming each of these four items will require (1) the prompt
localization of many more GRBs and XRFs; (2) the measurement of their
observed peak energy $\eop$, and fluence $\se$; (3) the determination
of their jet break time $\tjet$; and (4) the measurement of their
redshift $z$.  \hetetwo and \swiftns, working together, can do (1) very
well, enabling ground-based telescopes to do (4).  
\hetetwo is ideally suited to do (2), whereas the
\swift BAT (which has a relatively narrow energy band: 15 keV $< E <$
150 keV) cannot.  \swift XRT is ideally suited to do (3), whereas
\hetetwo cannot.  However, a scientific partnership between \hetetwo
and \swift can obtain the necessary precursor observations.

This scientific partnership works in two ways:  (1) a few bursts will
be observed by both satellites, since \hetetwo points anti-sun and
\swift is biasing its spacecraft pointing direction toward the anti-sun
direction as much as it can; and (2) \swift can slew to the 20-25
bursts per year that \hetetwo localizes and for which it measures
$\eop$ and $\se$.  The former has already happened for GRBs 050215b 
\citep{sakamoto2005b} and 050401 \citep{atteia2005,barbier2005}; the
latter has already happened for GRBs 050209 \citep{kawai2005} and
050408 \citep{sakamoto2005c,wells2005}. 

Of the 20-25 bursts per year that \hetetwo localizes, $\approx$ 7 are
GRBs, $\approx$ 7 are XRRs, and $\approx$ 9 are XRFs.  This implies
that, in three years of partnership with \swiftns, \hetetwo will
localize and determine $\eop$ and $\eiso$ for $\approx$ 21 GRBs and
XRRs, and $\approx$ 27 XRFs, giving a total of $\sim$ 70 bursts of
which it can reasonably be expected that $\approx$ 55  will have all
four quantities measured.  A few additional bursts localized by \swift
will have their $\eop$ and $\se$ measured by {\it Konus-Wind}.  Thus
precursor observations can provide $\approx$ 60 additional ``Gold''
bursts -- nearly doubling the size of the current sample of bursts with
known redshifts, quintupling the size of the current sample of ``Gold''
bursts, and increasing the number of XRFs with known redshifts from two
to 15-20 [XRFs 020903 \citep{soderberg2004} and 040701
\citep{kelson2003} are the only XRFs with accurately known redshifts; 
redshift constraints exist for two others: XRFs 020427
\citep{amati2004} and 030723 \citep{fynbo2004}].  The full sample of
$\approx$ 70 ``Gold'' bursts will enable a much more stringent test of
the degree to which GRBs obey the $\ep-\eiso-\tjet$ relation,
particularly XRFs.  In particular, confirmation that XRFs obey the
$\ep-\eiso-\tjet$ relation would make it possible to use them to
calibrate the relation (see below), and to increase the power of
the constraints on $w_0$ and $w_a$ (also see below).

\vspace{-0.1in}
\section{Error Budget}

{\it Statistical Errors:}  The primary sources of statistical errors
are the uncertainties with which the spectral parameters $\eop$, $\se$,
and the jet break time $\tjet$, can be determined.  In addition, there
are sample variances in the $\ep-\eiso$ and $\ep-\egamma$ relations of
sizes 0.2 and 0.1 dex, respectively.  Here we assume Gaussian
distributions perpendicular to both the best-fit $\ep-\eiso$ and the
best-fit $\ep-\egamma$ relations with widths $\sigma = 0.4$ and $\sigma
= 0.15$, respectively.  These widths accurately characterize the
current statistical plus sample variances in these distributions.  We
assume 20\% errors in both $\ep$ and $\tjet$, and 10\% errors in the
fluence $\se$.  The accuracies with which the  parameters (intercepts
$C$, slopes $s$, and intrinsic widths $\sigma$) of the $\ep$-$\eiso$
and $\ep$-$\egamma$ relations can be determined scale as
$1/\sqrt{N_{\rm GRB}}$, where $N_{\rm GRB}$ is the  number of bursts in
the sample used.  The uncertainties in the determination of $\omegam$,
$\omegal$, $w_0$, and $w_a$, therefore scale similarly.

The scientific partnership between \hetetwo and \swift discussed in the
previous section can produce significantly better determinations of the
spectral parameters $\ep$, $\eiso$, and $\egamma$ of the burst, and the
jet break time $\tjet$, especially for the important XRFs.  The mission
concept
that we propose would produce much better determinations of these
quantities.  Nevertheless, here we assume the previous accuracies with
which these quantities can be determined (i.e., 10\% errors in $\se$
and 20\% errors in $\eop$ and $\tjet$).  In this sense, the results that
we report in this white paper are very conservative.

{\it Systematic Errors:}
The use of GRBs to constrain the properties of dark energy is in its
infancy compared with many of the other approaches that are being used
and proposed.  We acknowledge that in this situation, the risk that
systematic errors may prevent this approach from reaching the accuracy
necessary to tightly constrain $w_0$ and especially, to remove the
degeneracy between $w_0$ and $w_a$ that exists if CMB priors are used
(as we do) may be greater than in some other methods.  However, there
are reasons to believe that GRBs may suffer less from systematic errors
than some other approaches.

\cite{fb2005} have discussed a number of possible sources of systematic
errors in using GRBs as standard candles.  \cite{gglf2005} and
\cite{fgga-v2005} have addressed many of these.  In addition, very
recent work by \cite{liang2005} has shown that the $\ep-\eiso-\egamma$
relation can be put on an entirely empirical footing by writing it in
terms of $\ep-\eiso-\tjet$.  This approach has the advantage of making
the relation explicitly model-independent, and eliminates the need to
marginalize over the density $n$ of the circumburst medium.

Perhaps the most worrying kind of systematic error for relative
standard candles is one that would mimic the effect of dark energy;
e.g., calibration errors that would produce a bump in the inferred
intrinsic brightness over the redshift range $0.6 \lesssim z \lesssim
1$ (i.e., $\Delta z/z \approx$ 40\%, corresponding to energy or
wavelength ranges $\Delta E/E \approx \Delta \lambda/\lambda \approx$
40\%).   GRBs may be less prone to this particular kind of systematic
error than other standard candles for several reasons.  First, the
determinations of $z$ and $\tjet$ are not subject to this kind of
systematic error.  Second, the spectra of GRBs are smoothly broken
power laws that span a broad energy range and that have a peak energy
in $\nu F_\nu$ that is broad.  Third, while the detector response
matrices (DRMs) of gamma-ray detectors (e.g, NaI) have an absorption
feature at $E \approx 37$ kev, the DRMs are not diagonal: because of
Compton scattering and other effects, a beam of mono-energetic photons
produces counts in a range of energy loss bins, which smooths the
detector response as a function of energy.  These properties of GRB
spectra and detectors lessens the probability that systematic errors of
the kind that would mimic the effect of dark energy can occur.

\begin{figure*}
\centerline{
\includegraphics[scale=0.7,clip=]{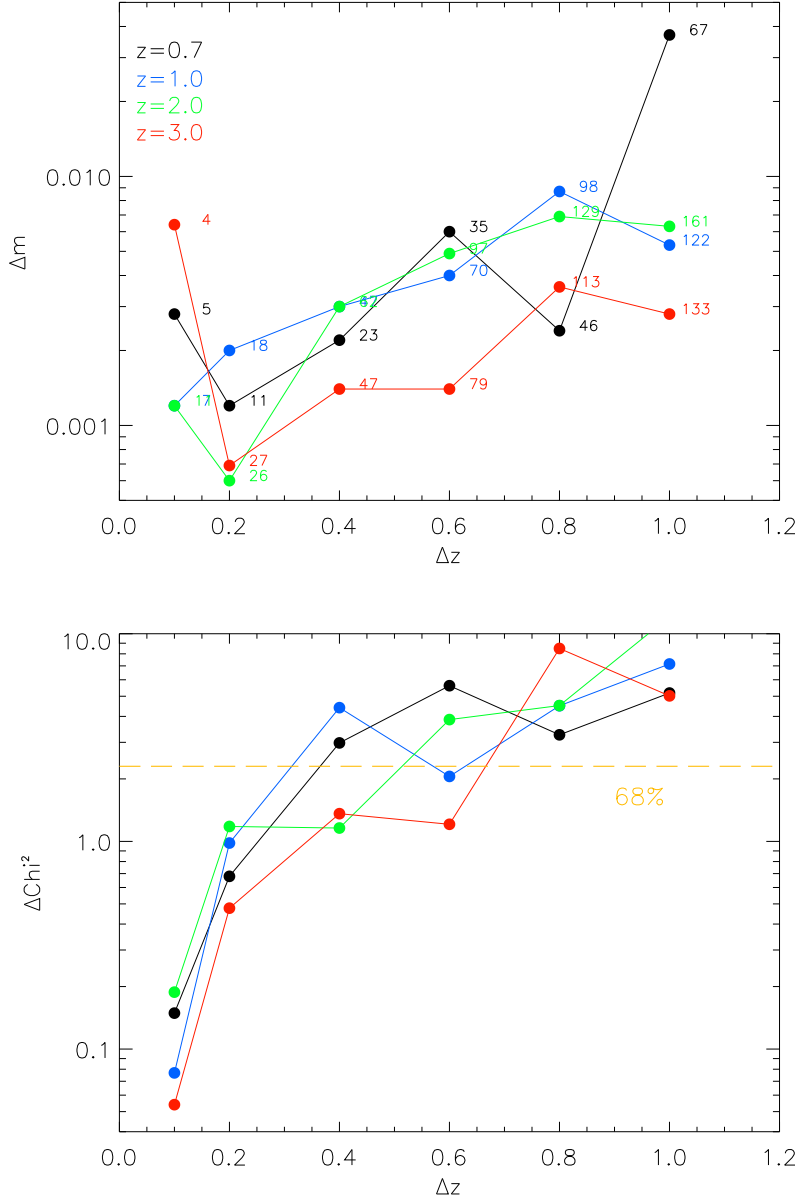}}
\caption{Illustration of the self--calibration method for the
$\ep-\eiso$ relation.  For different redshift ``slices''  $\Delta z$,
centered on redshifts $z =$ 0.7, 1.0, 2.0, and 3.0, the following
quantities are shown:  Top panel:  Maximum variation $\delta m$ in the
slope $m$ of the $E_{\rm peak}-E_{\gamma}$ correlation, obtained from 
surveying large ranges in the cosmological parameters $\omegam$ and
$\omegal$.  The numbers next to each point are the number of bursts in
that particular redshift slice.  Bottom panel:  Maximum variation in
$\chi^{2}$ obtained from  surveying large ranges in the cosmological
parameters $\omegam$ and $\omegal$.  Also shown is the 68\% confidence
level.  A value of $\Delta\chi^{2}$ below this level ensures that the
$\ep-\eiso$ relation found using that redshift range (i.e., that size of 
$\Delta z$) is cosmology independent.}
\label{fig:self-cal}
\end{figure*}

\begin{figure*}
\centerline{
\includegraphics[scale=0.9,clip=]{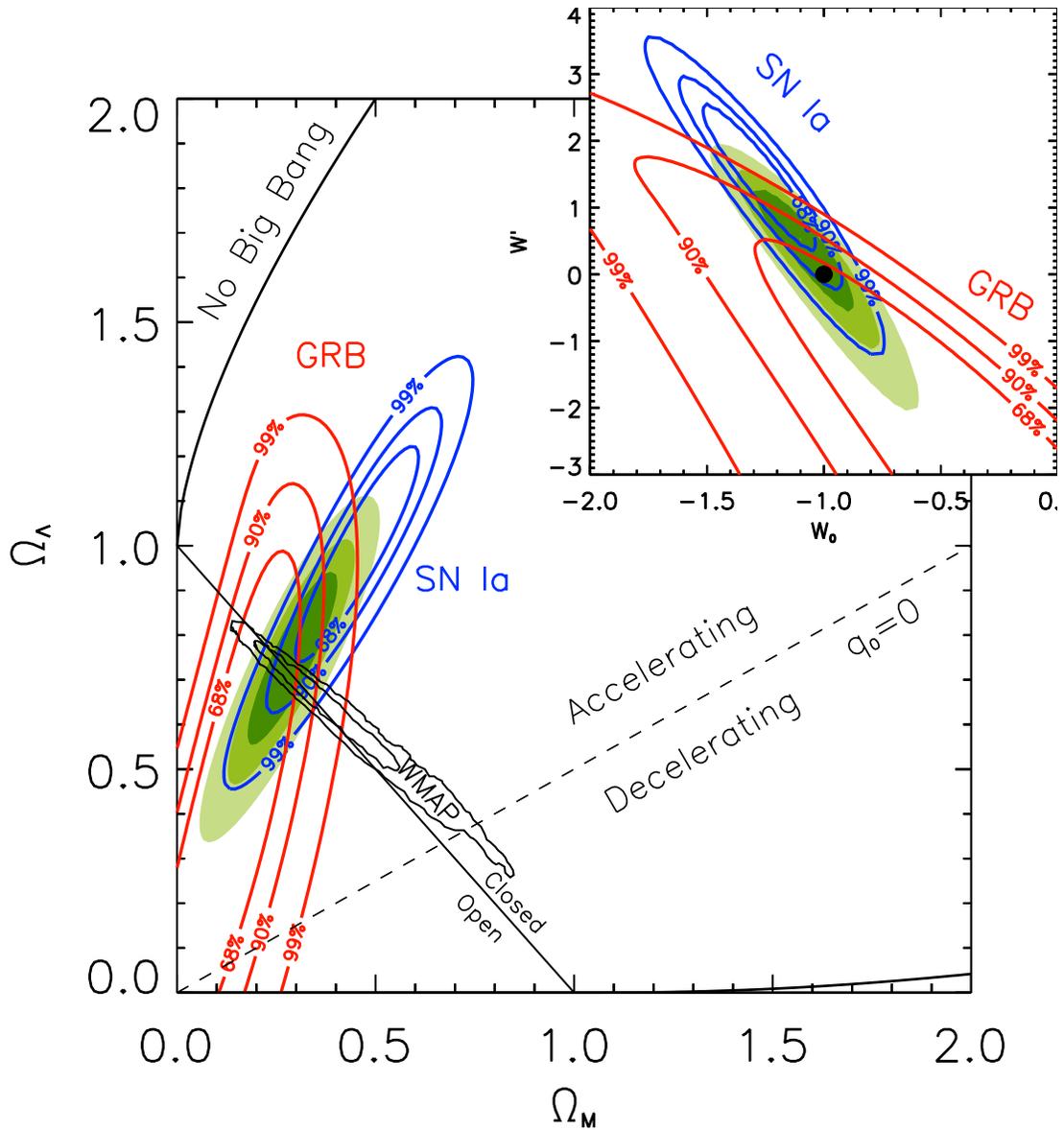}}
\caption{Example of how GRBs can contribute to the determination of the
cosmological parameters once the $E_\gamma - E_{\rm peak}$ correlation
can be determined in a cosmology-independent way [i.e., accumulating
sufficient bursts at small redshifts or in a small redshift range (e.g.
around $z = 2$)].  We show the contours in both the
($\omegam,\omegal$)- plane ($\omegal=\Omega_{\Lambda}$ in main 
figure) and in the $w_0-w^\prime$
plane (insert, a flat cosmology with $\omegam=0.27$ is assumed). Lines
and colors are as in Fig. 2 and Fig. 3.  (Figure from \citet{gglf2004}).}
\label{fig:ghirlanda_2_fig4}
\end{figure*}

{\it Self-Calibration:} The $\ep-\eiso-\tjet$ relation is cosmology
dependent because $\eiso$ depends on cosmology \citep{gglf2004}. 
However, the fact that the 15 bursts in the current ``Gold'' sample of
GRBs span the broad redshift range $z$ = 0.17 - 3.2 is evidence that
the relation does not evolve with redshift; i.e., it is independent of
$z$.  Assuming that the $\ep-\eiso-\tjet$ relation is independent of
$z$, a self-calibration  approach can be used to eliminate the
dependence of its slope on cosmology.  In this approach, either a
sample of XRFs at low redshifts or a sample of GRBs in a narrow
redshift range at high  redshifts can be used to determine the
relation, independent of cosmology.  Figure \ref{fig:self-cal}
illustrates the latter approach.  It shows that, using this method, the
slope of the relation can be calibrated to $<$ 1\% accuracy,
independent of cosmology.  Comparisons of the parameters that define
the relation for different redshift intervals, and for bursts at low
redshift, can be used to check that the slope of the relation is indeed
independent of cosmology.  Figure \ref{fig:ghirlanda_2_fig4}
illustrates the much smaller sizes of the confidence regions that
result when the slope of the $\ep-\egamma$ relation is known,
independent of cosmology.

Lastly, the mission concept that we are pro- posing would provide a data
base of\ $>$ 800 bursts with well-determined spectral parameters, \, jet
break times, and redshifts that will be an order of magnitude larger
than the sample of such bursts that will be produced by the scientific
partnership between \hetetwo and \swift.  This large sample of ``Gold''
bursts will enable a search for the best empirical relationship among
the spectral parameters, the jet break time of the afterglow, and the
redshift thus tightening the constraints on the properties of dark
energy.  It will also provide an unprecedented data base for gaining a
new understanding of GRB jets -- and therefore minimizing -- possible
sources of systematic error.

\vspace{-0.1in}
\section{Constraints on Nature of Dark Energy}

Scaling from the burst rate seen by BATSE, \bsax, and \hetetwo, the
mission concept we are proposing would localize and obtain the spectra
for $\approx$ 800 GRBs per year, or 1600 during a 2-year mission, of
which 200 per year, or 400 during a 2-year mission, would be XRFs.  The
percentage of \hetetwo bursts for which redshifts have been obtained
during the past three years is $\sim$ 30\%; however, most of these
bursts had $7\arcmin-10\arcmin$ WXM localizations, requiring relatively
wide-field optical cameras to identify the optical afterglow, a
prerequisite to determining the redshift of the burst from absorption
lines in the afterglow spectrum.  In contrast, the mission concept that
we are proposing would obtain and disseminate arcsecond burst
localizations in real time.   We therefore estimate that redshifts will
be obtained for 50\% of the bursts that the mission localizes, a
percentage that we regard as quite conservative.  This gives 400 bursts
per year, or 800 bursts in a 2-year mission, with well-measured values
of $\eop$, $\se$, $\tjet$, and $z$.

\begin{figure*}
\centerline{
\includegraphics[scale=0.9,clip=]{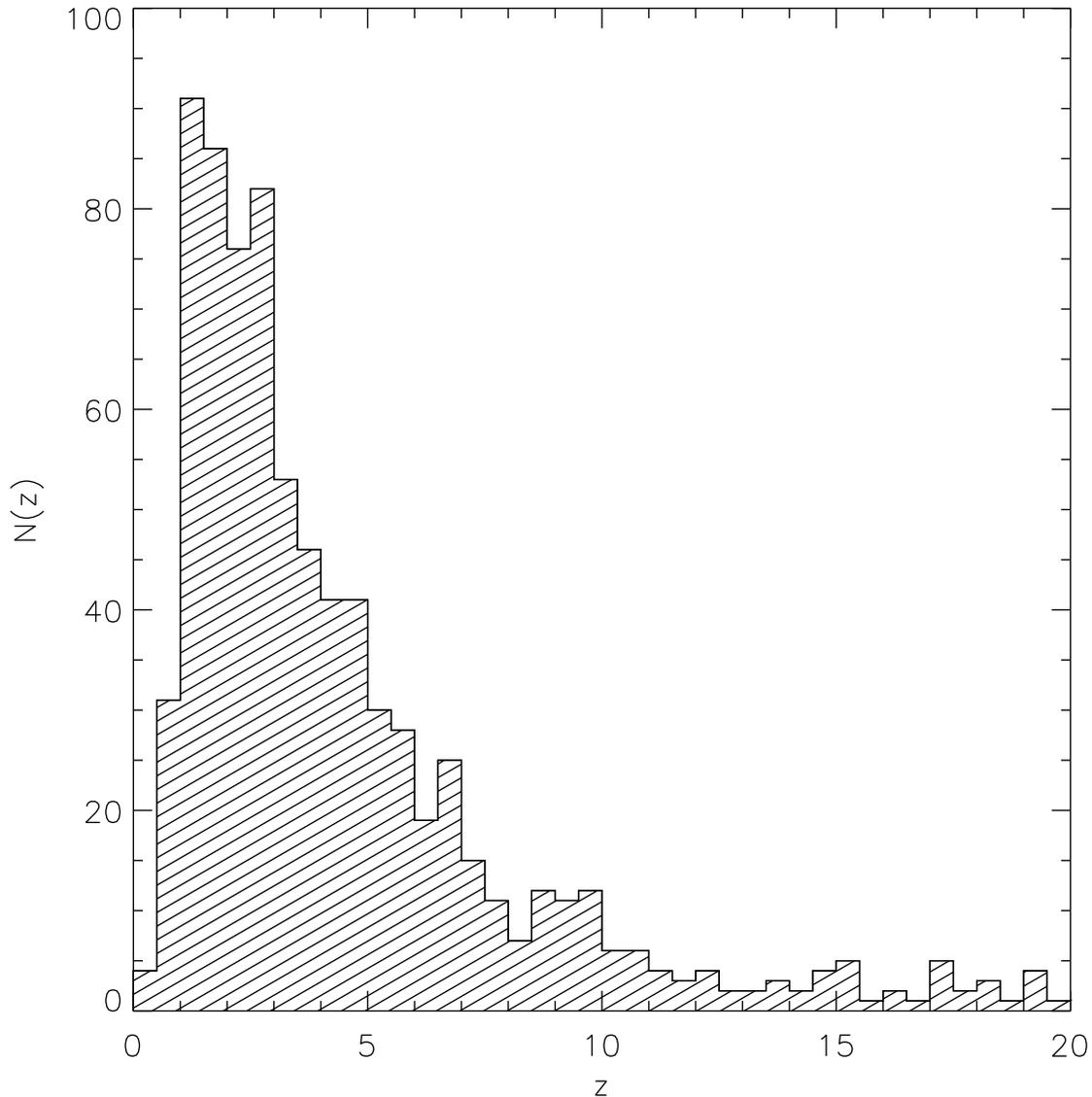}}
\caption{Distribution of GRBs as a function of redshift in our
simulated catalog of 800 bursts.  The distribution of GRBs is
proportional to the phenomenological expression for the star formation
rate (SFR) suggested by \cite{rowan-robinson2001}, adopting the
parameter values $P = 1.2$ and $Q = 5.4$ (which provides a good fit to
existing data on the SFR as a function of redshift).  With this choice
of the SFR and the nominal mission design, the catalog contains 35 GRBs
at $z < 1$.  However, XRFs have redshifts $z \lesssim 0.5$, and by
adjusting the design of the proposed mission concept, the catalog could
contain as many as 200-300 XRFs at very low redshifts.
}
\label{fig:grb_rate}
\end{figure*}

We compute the constraints that GRBs can place on the properties of
dark energy as follows.  First, we generate a catalog of 800 GRBs with
spectral properties matching the distributions of the spectral
properties of GRBs seen by \bsax and \hetetwo and a redshift
distribution matching the inferred star formation rate as a function of
redshift (see Figure \ref{fig:grb_rate}).  We parameterize the dark
energy equation of state $w(z)$ as a function of redshift by  the
phenomenological expression \citep{linder2003} $w(z) = w_0 + (1-a)
w_a$, where $a = 1/(1+z)$ is the scale factor of the universe.  In
constructing the catalog of simulated bursts, we assume $\omegam =
0.30$ and $\omegal = 0.7$, $w_0 = -1$, and $w_a =0$ (i.e., a spatially
flat, vanilla $\Lambda$CDM cosmology).

\begin{figure*}
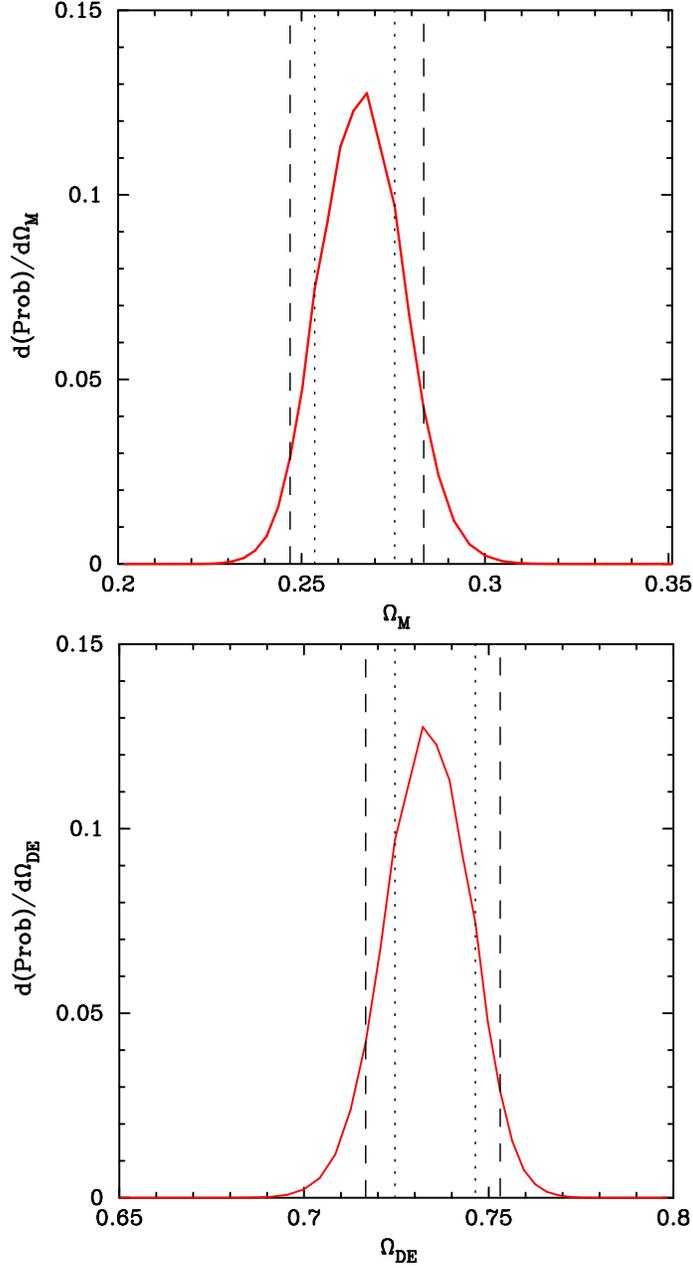

\centerline{
\includegraphics[scale=0.5,clip=]{figures/omegam_1d_posterior.ps}}
\centerline{
\includegraphics[scale=0.5]{figures/omegal_1d_posterior.ps}}
\caption{Constraints on $\omegam$ and $\omegal$ for 800 GRBs, using CMB
priors (the only free parameter is then $H_0$).  Top panel: posterior
probability distribution for $\omegam$.  Bottom panel: posterior
probability distribution for $\omegal$.  The dotted lines show the 68\%
credible regions; the dashed lines show the 90\% credible regions.}
\label{fig:omega_2yr}
\end{figure*}

\begin{figure*}
\centerline{
\includegraphics[scale=0.6,clip=]{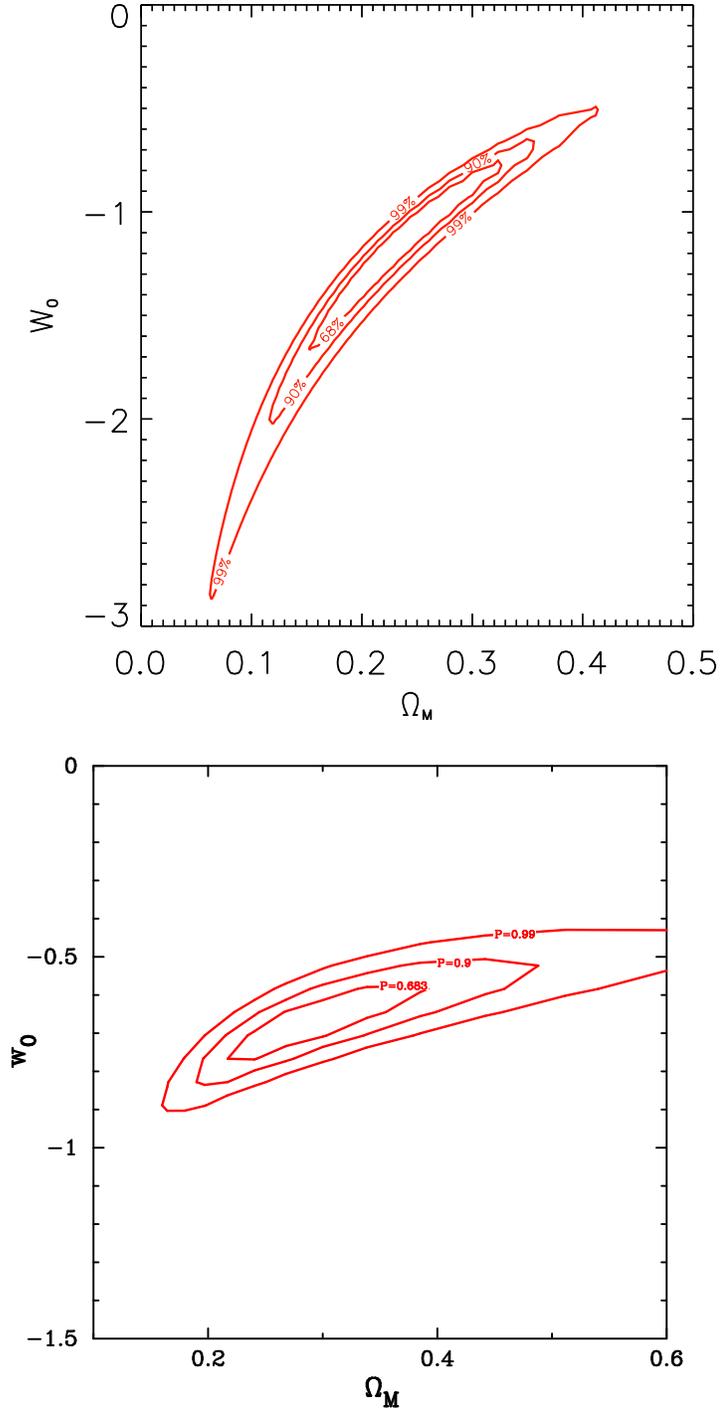}}
\centerline{
\includegraphics[scale=0.5]{figures/omega_w0_hprior.ps}}
\caption{Top panel: $\omegam$ versus $w_0$ for 800 GRBs and using CMB
priors.  Bottom panel: the same, except that we also use the prior $H_0
= 68 \pm 8$ km s$^{-1}$ Mpc$^{-1}$ after the HST key project
\citep{freedman2001}.}
\label{fig:w_omega_2yr}
\end{figure*}

\begin{figure*}
\centerline{
\includegraphics[scale=1.1,clip=]{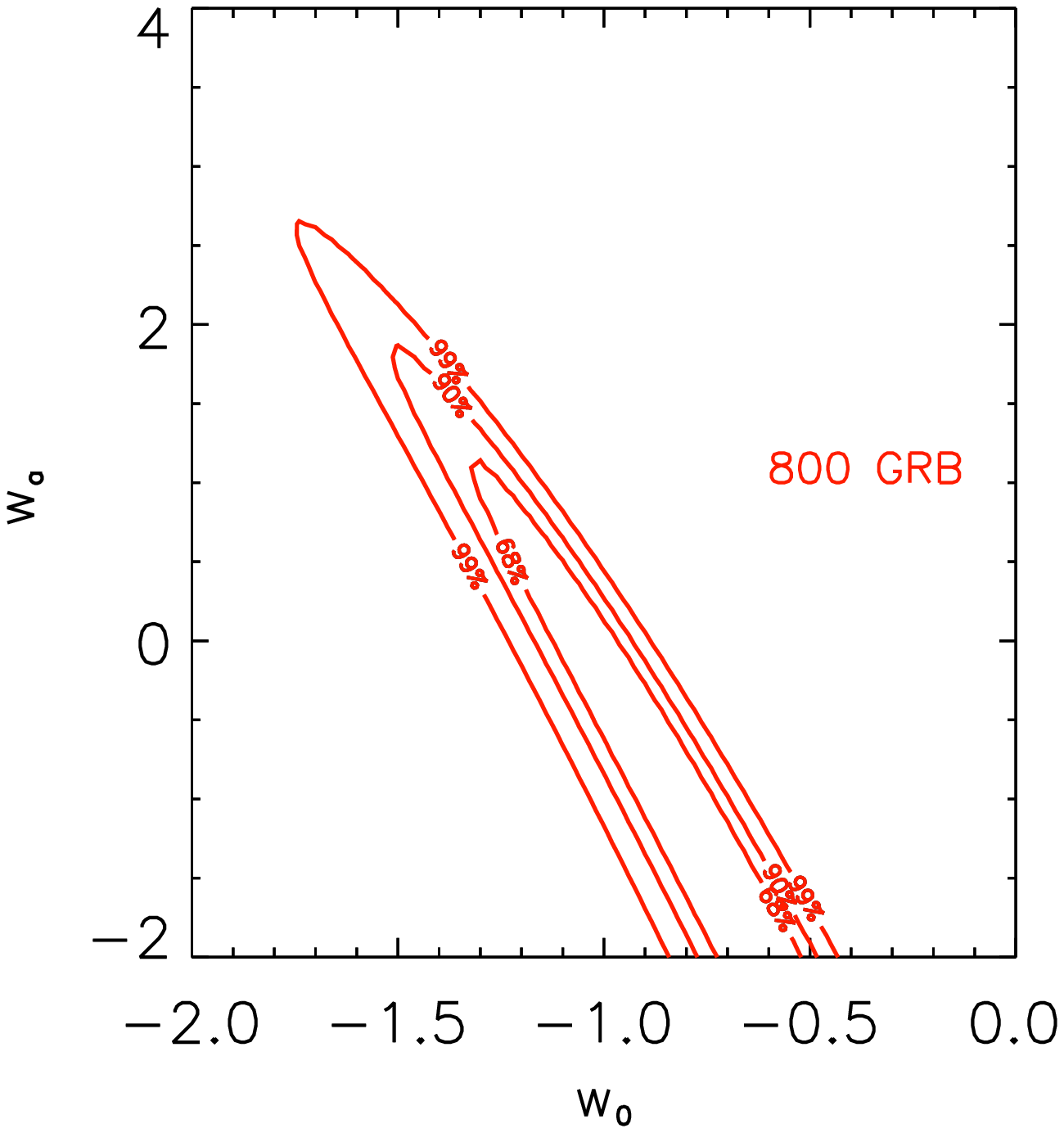}}
\caption{Constraints on $w_0$ and $w_a$ for 800 GRBs, using CMB priors
and the prior $H_0 = 68$ km s$^{-1}$ Mpc$^{-1}$ after the HST key 
project \citep{freedman2001}.  The last prior is a good approximation,
since the constraints derived on $w_0$ and $w_a$ are much larger than
this.  However, using the correct prior, $H_0 =  68 \pm 8$ km s$^{-1}$
Mpc$^{-1}$ [after the HST key  project \citep{freedman2001}], would
widen the contours in the direction perpendicular to their long axes.}
\label{fig:w0_wa_2yr}
\end{figure*}

We determine the best-fit cosmology and 68\%, 90\%, and 99\% confidence
regions by comparing the luminosity distance $d_L(z)$ for various
cosmologies with the distribution of $d^{\rm GRB}_L$ values for the 800
GRBs in our simulated burst sample, using $\chi^2$ as the statistic. 
In doing so, we employ two CMB priors:  $\omegam + \omegal = 1$ (a flat
universe) and $\omegam h^2 = 0.14$.  Because Planck is expected to
determine the value of the latter to  $<$ 1\%, and to simplify the
computations, we assume that both priors have zero uncertainty.  Given
these priors, $d_L(z)$ is a function only of $H_0$, $w_0$, and $w_a$.
We consider three cases:  

\begin{enumerate}

%\vspace{-0.1in}
\item
We first assume that $w_0 = -1$ and $w_a = 0$.  The only free parameter
is then $H_0$.  Figure \ref{fig:omega_2yr} shows the resulting
posterior  probability distributions for $\omegam$ and $\omegal$.

%\vspace{-0.1in}
\item
We assume that $w_a = 0$.  The free parameters are then $H_0$ and
$w_0$.  Figure \ref{fig:w_omega_2yr} (top panel) shows the resulting
68\%, 90\%, and 99\% confidence regions in the ($\omegam,w_0$)-plane. 
Figure \ref{fig:w_omega_2yr} (bottom panel) is the same, except that we
also employ the prior $H_0 = 68 \pm 8$ km s$^{-1}$ Mpc$^{-1}$ after the
HST key project \citep{freedman2001}, but adopting a best-fit value for $H_0$
that is consistent with the values of $\omegam$ and $\omegal$ that we
used in constructing our catalog of simulated bursts.

%\vspace{-0.1in}
\item
We make no assumptions about $w_0$ and $w_a$.  We again employ the
prior on $H_0$ from the HST key project, but to simplify the
computations we assume that it has zero uncertainty (this is a good
approximation, since the constraints derived on $w_0$ and $w_a$
are much larger than this.)  The free parameters are then $w_0$ and
$w_a$.  Figure \ref{fig:w0_wa_2yr} shows the resulting 68\%, 90\%, and
99\%  confidence regions in the ($w_0$,$w_a$)-plane.

%\vspace{-0.1in}
\end{enumerate}

\noindent
Table 1 gives the 1-$\sigma$ uncertainties in the dark energy
parameters for each of the above cases, marginalized over the other
parameters.  The assumed values of the parameters lie within the 68\% 
confidence regions but not at their centers (unlike Fisher  matrices)
because we use a simulated catalog of GRBs.

\begin{figure*}
\centerline{
\includegraphics[scale=1.2,clip=]{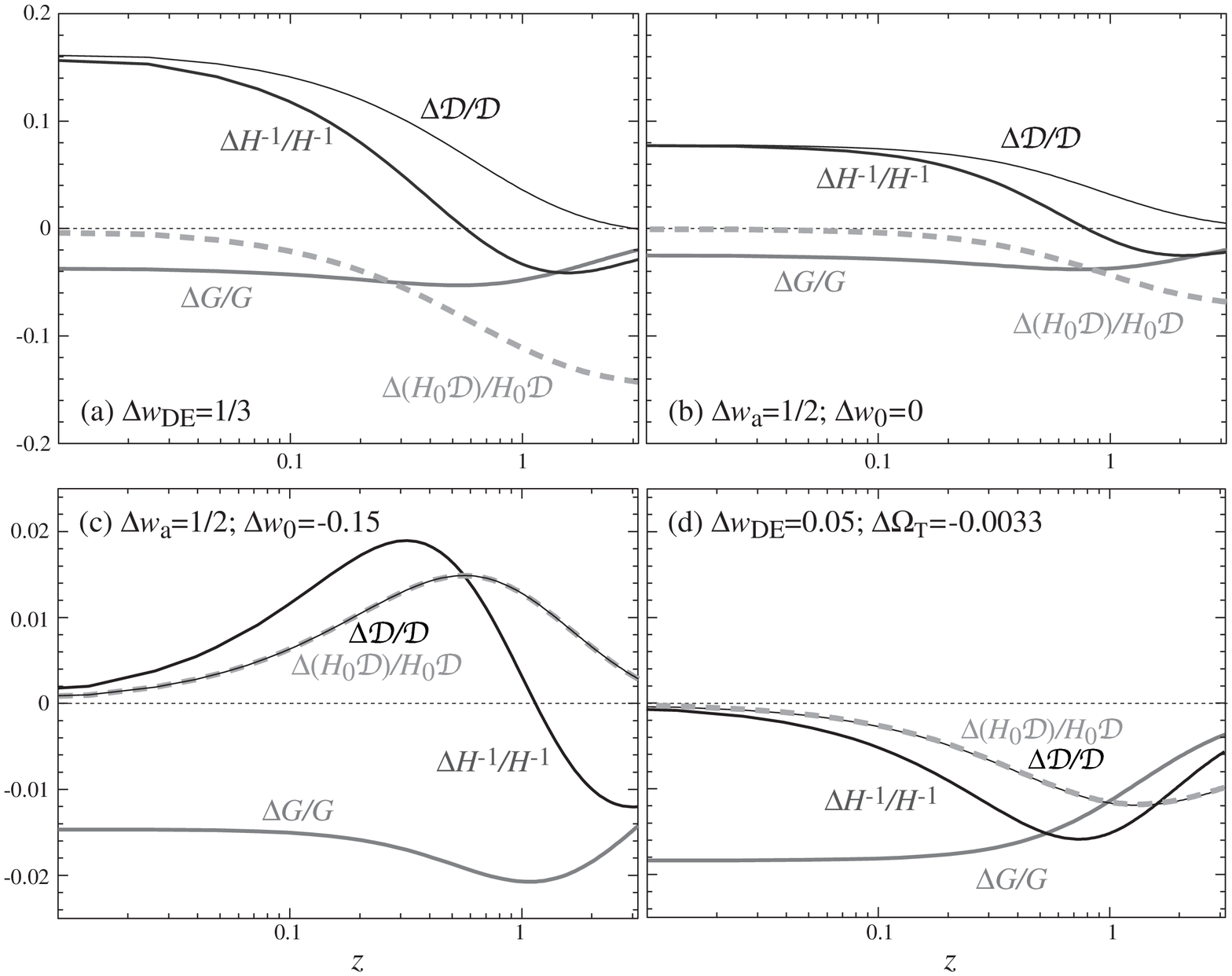}}
\caption{Sensitivity of determination of $w_0$, (assuming $w_a = 0$) as
a function of redshift for various quantities.  Both GRBs and Type Ia
SNe are ``relative standard candles,'' in the sense that their
intrinsic luminosities are not determined; instead, their relative
brightnesses are compared at different redshifts.  Many GRBs occur and
can be observed at high redshift.  Using CMB priors then provides an
absolute calibration \citep{hu2005}.  In this case, one is measuring
$\Delta D/D$ and the strongest constraints on whether $w_0 = -1$ come
from GRBs observed at $z \lesssim 0.3$.  This is opposite to the case
for Type Ia SNe, since many Type Ia SNe have been observed at $z
\lesssim 0.3$; then $\Delta H_0 D/H_0 D \rightarrow \Delta H_0/H_0$,
and the strongest constraints on whether $w_0 = -1$ come from Type Ia
SNe observed at redshifts $z > 1$.  From \cite{hu2005}.}
\label{fig:hu}
\end{figure*}

GRBs, like Type Ia SNe, are ``relative'' standard candles, in the sense
that their intrinsic luminosities are not determined; instead, their
relative brightnesses are compared at different redshifts.  Many GRBs
occur and can be observed at high redshift.  Using CMB priors then
provides an absolute calibration \citep{hu2005}.  In this case, one is
measuring $\Delta D/D$ and the most severe constraint on whether $w_0 =
-1$ comes from GRBs observed at $z \lesssim 0.3$.  This is opposite to
the case for Type Ia SNe, since many Type Ia SNe occur at $z \lesssim
0.3$; then $\Delta H_0 D/H_0 D \rightarrow \Delta H_0/H_0$, and the
most severe constraint on whe- ther $w_0 = -1$ comes from Type Ia SNe
observed at redshifts $z > 1$ (see Figure \ref{fig:hu}).  

Therefore, in order to exploit the power that comes from using CMB
priors, a sufficient number of bursts must be observed at very low
redshifts.  If the precursor observations described above confirm that
XRFs satisfy the $\ep-\eiso-\tjet$ relation, XRFs can provide the
population of relative standard candles at $z \sim 0$ that is needed: 
XRFs have redshifts $z \lesssim 0.5$ and by adjusting the design of the
proposed mission concept, as many as 200-300 XRFs could be obtained at
very low redshifts.  Alternatively, the complementary redshift ranges
of Type Ia SNe ($0 \lesssim z \lesssim 1.7$) and classical GRBs ($0.5
\lesssim z \lesssim 10$) could be exploited to provide a combined
population of relative standard candles that spans the entire redshift
range $0 \lesssim z \lesssim 10$, which could provide more severe
constraints on the properties of dark energy than could either
population alone.

\vspace{-0.1in}
\section{Proposed Mission Concept}

The mission concept (see Appendix A) is optimized to use GRBs as
``standard candles'' to determine the properties of Dark Energy and probe
the early universe rather than increasing our understanding of GRBs and
their precursors.  In this sense, it is very different than previous
GRB missions, including {\it Beppo}SAX, {\it HETE-2}, and {\it Swift}, 
although it will provide a great deal of information about the prompt
and afterglow emission of a large number of bursts.  The mission
concept that we describe also builds on the ``lessons learned'' from
previous GRB missions, including {\it Beppo}SAX, \hetetwo, and {\it
Swift}.  In particular, it exploits the very recent discoveries that
tight relations exist between $\eiso$ and $\ep$, and between $\egamma$
and $\ep$, and it combines the capabilities of all of these missions in
a way that maximizes the scientific return of the mission for the
purpose of determining the properties of dark energy.

As described more fully in Appendix A, the mission concept calls for
three small spacecraft operating in near-proximity at the sun-earth $L_2$
point for two years with the following roles: (1) One Prompt Satellite
(a ``mini spacecraft,'' $\sim$200 kg) to detect and fully characterize the
prompt spectra of GRBs and XRFs over the 1-1000 keV band, as well as to
establish their location to $\sim$arcsecond accuracy. (2) Two Afterglow
Satellites (``micro spacecraft,'' $\sim$50 kg each) to measure the burst
X-ray afterglow and to determine the time of the jet break. In
addition, as described in Appendix B, the mission concept calls for:
(3) A network of six fully-automated, two meter-class ground-based
telescopes equipped with integral-field spectrometers (IFS) that will be
dedicated to establishing redshifts for a large fraction ($\sim$50\%)
of the localized GRBs and XRFs.

\vspace{-0.1in}
\section{Other Issues}

{\it Risks and Strengths:}  This mission concept is based on a low-risk 
approach
that relies on flight-proven technologies and the heritage of {\it HETE-2}. 
Redundancy is provided by a cluster of silicon X-ray CCD detectors and a set of
NaI detectors on one satellite, and by two focusing X-ray
telescope arrays, each on a separate satellite.  The number of GRBs observed
is large; the contribution of statistical errors to the uncertainties
with which the dark energy parameters can be determined will be
modest.  The level at which systematic errors will dominate is not
known, and represents a risk, but precursor observations can
substantially reduce this risk.  In addition, the properties of GRB
spectra and gamma-ray detectors lessen the probability that systematic
errors of the kind that would mimic the effect of dark energy can
occur.

{\it Technology Readiness:}  There are no complicated requirements on
the satellites or the operations.  The Prompt Satellite is of
modest size; the X-ray CCD detectors are similar to those used in the
Soft X-ray Camera (SXC) on \hetetwo and the focal plane of {\it
Chandra}, and the NaI detectors are nearly identical to those that have
flown on numerous GRB missions over the past 30 years--most recently
Fregate on {\it HETE-2}. The two X-Ray Afterglow Satellites are of very
small size; the X-ray mirrors are of short focal length and are small in
diameter. There X-ray CCD detectors are even smaller than those that were
flown on {\it ASCA}. The required sensitivity
for this mission concept is achieved primarily by greatly increasing
the number of detector modules, so as to view a much larger solid angle
($\sim 10$ times that of \hetetwo and {\it Swift}) and by operating at
$L_2$, which permits a much greater observing efficiency ($\sim 3$ times
that of {\it HETE-2}). The dedicated ground segment of the mission
relies on purchasing six commercially-available two-meter class telescopes
and equipping them with efficient, state-of-the-art integral field
spectrometers. Thus, the mission concept relies on flight-proven
technologies and established ground-based telescope systems, and is
thus ``ready to go.''

{\it Relationship to JDEM and LSST:}  The GRB Dark Energy mission is
not a precursor to JDEM or LSST.  Indeed, the significantly different
parameter degeneracies of the constraints on the parameters that
characterize Dark Energy that will come from the GRB Dark Energy
mission make it complementary to JDEM and LSST, as well as
to {\it Planck}~and any X-ray survey of galaxy clusters.

{\it Access to Facilities:}  All facilities that are needed for the
baseline mission have been included as part of the mission.
Ground-based optical follow-up observations using existing large-aperture
($>6$ meter)
telescopes to determine the redshifts of some fainter bursts (e.g.,
some XRFs) would increase the number of bursts beyond that assumed in
this white paper.

{\it Timeline:}  The GRB dark energy mission discussed in this white
paper is not currently funded.  The baseline mission fits within the
resources (time, mass, volume, and cost) of a NASA Medium Explorer
mission.  It could therefore be proposed for next Midex AO and flown 4
years later (e.g., in $\sim$ 2011).

\vfill
\eject

\clearpage

\begin{table*}[h]
\centerline{Table 1. Dark energy parameter constraints from 800
GRBs.$^{a}$\phantom{xxxxxxxxxxxxxxxxxxx}}
\begin{tabular}{cccc}
\hline\hline
Priors$^{a}$  & $\sigma(\omegal)^{b}$ & $\sigma(w_{0})^{b}$ &
$\sigma(w_{\rm a})^{b}$ \\
\hline
$w_{0}=-1$, $w_{\rm a}=0$ & $-0.0076$, $+0.0142$ & --- & --- \\
$h=0.68 \pm 0.08$, $w_{\rm a}=0$ & $\pm 0.070$ & $\pm 0.061$ & --- \\
$\omegam \equiv 0.3$ & --- & $-0.2$, $+0.32$ & $-1.12$, $+0.63$ \\
\hline
\end{tabular}

$^{a}$All results assume CMB priors (i.e., $\omegam + \omegal = 1$
and $\omegam h^2 = 0.14$), \\
{\phantom{$^{a}$}}in addition to the priors listed.

$^{b}$All confidence regions are 68.3\% CL.
\label{table:constraints}
\end{table*}
\clearpage

\clearpage

\clearpage

\clearpage

\clearpage

\clearpage

\clearpage

\clearpage

\clearpage

\clearpage

\clearpage

\clearpage

\vfill
\clearpage
\eject
~
\vfill
\eject
\clearpage

\twocolumn
\begin{figure*}[ht]
\centerline{{\Large \bf Appendices: Instrumentation Concept for}}
\centerline{{\Large \bf GRB Dark Energy Investigation}}
\end{figure*}

\vfill
\eject
\clearpage
~
\vfill
\eject
\clearpage

\setcounter{figure}{0}
\renewcommand{\thefigure}{A.\arabic{figure}}

\begin{figure*}[htb]
\centerline{
\includegraphics[width=5in]{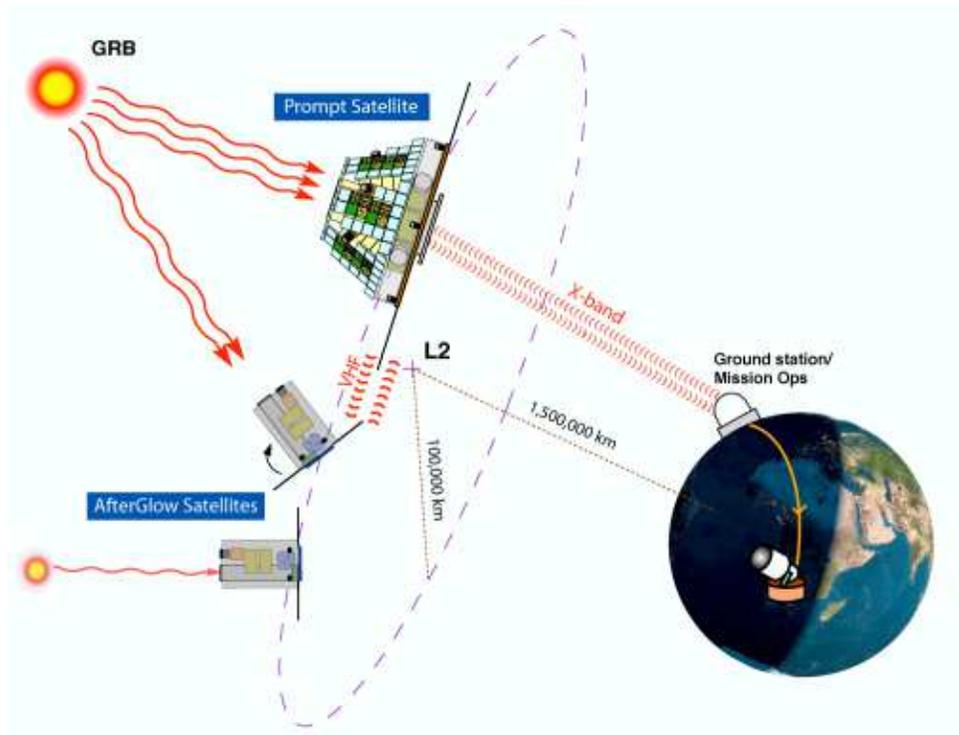}}
\caption{Schematic diagram of spacecraft locations and operation.}
\label{fig:schemdia3}
\end{figure*}

%\vfill
%\eject
%\clearpage

\centerline{\bf Introduction}

As outlined in the text,
for each of the 800 GRBs in the sample, we need to measure the spectral
peak, $\ep$; the isotropic energy fluence,  
$\eiso$; the jet break time,  $\tjet$;
and the redshift, $z$.
In the concept we describe below, sensitive silicon and sodium iodide 
detectors mounted on a ``Prompt Satellite'' will establish 
$\ep$ and $\eiso$, as well as burst location.
A dedicated optical ground network will measure the redshift, $z$, as 
described in Appendix B. Compact X-ray Afterglow
Telescopes (XATs), flying on two ``Afterglow Satellites'' will 
measure $\tjet$.  Figure \ref{fig:schemdia3} depicts the interaction
between the assembled assets for the mission concept.

The key to measuring $\tjet$ and $z$ is rapid
and precise determination of astrometric positions, allowing the optical
ground network and XATs to acquire the afterglow while it is still 
bright. Soft X-ray Camera (SXC) modules with
5 $\mu$m pixels on the Prompt Satellite designed for precision astrometry 
should be
capable of locating bursts to an accuracy of 1\arcsec-3\arcsec~in real time. This
is sufficiently accurate for immediate ground-based IFS spectroscopy
and for pointing a satellite-borne XAT with a narrow field of view.

The {\it HETE-2}~SXC has selected a much higher yield of bright afterglows than
other  GRB locating instruments. In addition, the distribution of
redshifts in this sample is very flat out to $z>3$ \citep{berger05}.
This is exactly the kind of sample that is required for
a dark energy investigation.

Relative to the \hetetwo SXC, the instrument
concept we propose will cover a much greater solid angle with much higher
operational efficiency. This broad coverage is the only way to acquire
large numbers of bright bursts. The \hetetwo
SXC localizes $\sim 12$ bright bursts/year
in a $\sim 0.6$ sr field of view with an operational efficiency of $\sim$
20\%. An
array of \hetetwo SXCs with hemispherical coverage and 80\% operational
efficiency would therefore localize $\sim 500$ bright bursts/year. This is a
very conservative lower limit on the burst rate for this instrument concept,
which will have $\sim 30$ times the detector area per solid angle and will
therefore be far more sensitive than {\it HETE-2}.  If the burst rate scales
as $A^{1./3}$ \citep{doty04}, this would imply that we might anticipate
$\sim$ 1600 events per year.  We adopt a conservative value of 800 per year.

The Prompt Satellite will also carry a large array of sodium iodide (NaI) 
scintillators to accurately measure burst spectra and fluences in the 5-1000 
keV range. The technology for such detectors is well-established, based on 
their use in GRB satellites from the 1960s until the present-- most recently 
in the Fregate instrument on {\it HETE-2}. Compared to Fregate, the proposed 
NaI array will provide 75 times larger area, 3 times greater solid angle, 
and 3 times greater observing time per year. Based on the Fregate rate 
of $\sim 90$ bursts/year, we conservatively estimate that the Prompt 
Satellite will detect $\sim  1000$ bright GRBs and XRFs per year--from
which we can extract high S/N spectra for accurately measuring 
$\alpha$, $\beta$, $\ep$, and the bolometric fluence over 
the 5-1000 keV range. The SXC silicon detectors will provide complementary 
spectra in the 1-10 keV range with the required sensitivity.
In the following two appendices, we provide additional details on the space and
ground segments of the mission concept, respectively.

\appendix
\setcounter{table}{0}
\renewcommand{\thetable}{A.\arabic{table}}
\twocolumn
\section{Instrument Concept for Space-based Segment--GRB Satellites at L$_2$}

\subsection{Prompt Satellite}

\begin{figure*}[htb]
\centerline{
\includegraphics[width=5in]{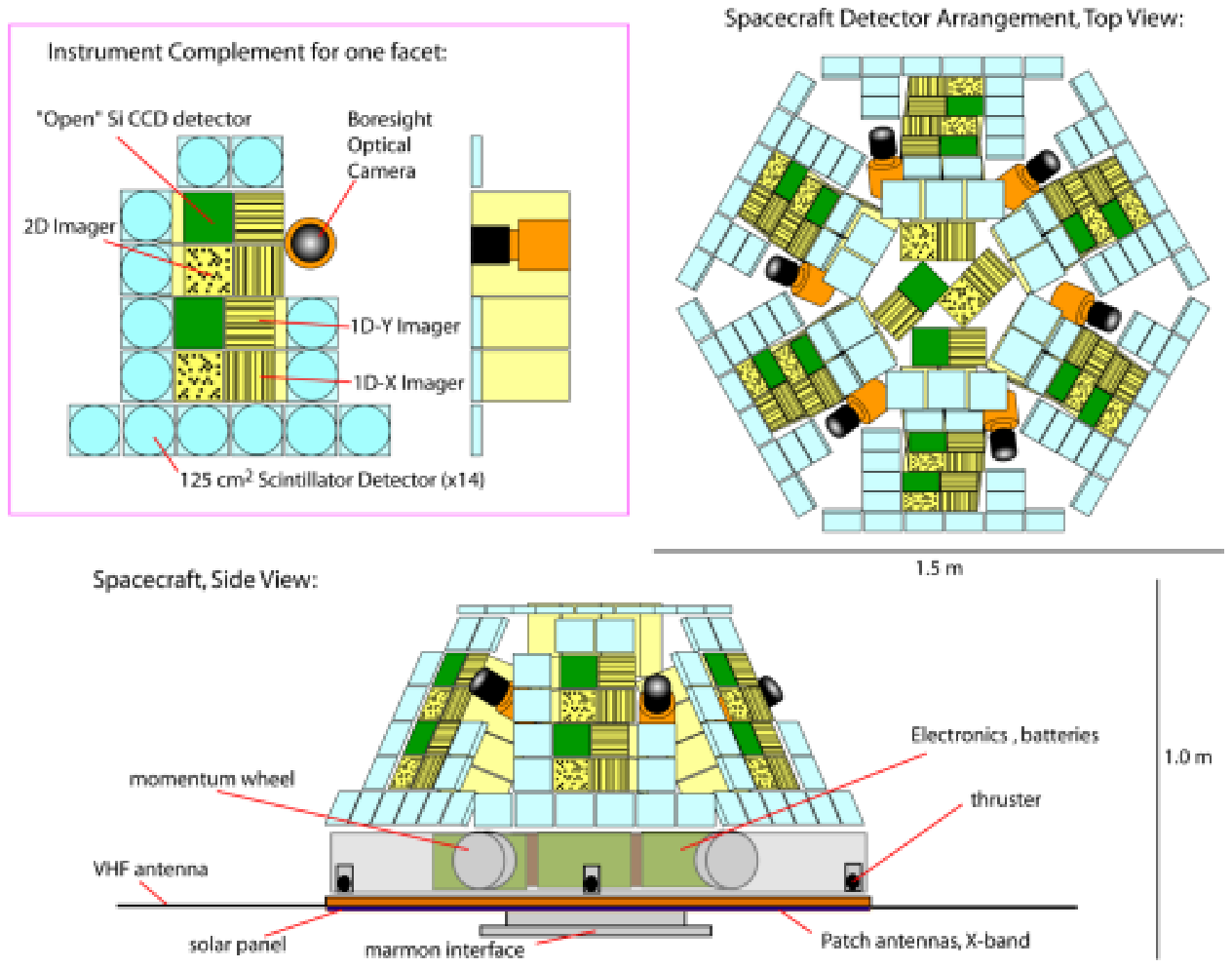}}
\caption{Schematic diagram of the GRB Prompt Satellite.}
\label{fig:schm}
\end{figure*}

The Prompt Satellite is depicted in Figure \ref{fig:schm}.  From its location
at L$_2$, the detectors on its seven facets view the sky continuously.  Each
facet contains 14 wide-angle ($\sim 140^{\circ}$ FOV) NaI detectors
(1750 cm$^2$ area per facet), covering the 5-1000 keV energy range.  Because
the fields-of-view of the NaI detectors on adjacent facets overlap, it is
possible to establish crude ($\sim 6^{\circ}$) localizations from detector
counting
rate ratios (as was done for the {\it BATSE}~mission).  Additional details
on the NaI detectors are given in Table \ref{table:scint_params}.

\begin{table*}[htb]
\begin{center}
\caption{Scintillation Detectors for Prompt Satellite}
\begin{tabular}{ll}\hline\hline
Energy Range & 5 - 1000 keV \\
Field of View & 7 sr total \\
Number of Modules & $14 \times 7  = 98$ \\
Scintillator &    NaI(Tl) \\
Readout System & Photomultipliers (PMTs) \\
Sensitive Area & 12,250 cm$^2$ ($14\times 7= 98$ units @ 125 cm$^2$) \\
Source Localization & 6$^{\circ}$ (by projected area method) \\
Flux Sensitivity (10$\sigma$, 8-1000 keV) & $10^{-8}$ ergs
cm$^{-2}$  s$^{-1}$ ($3\times$ better than BATSE \& Fregate) \\
Time Resolution &     10  $\mu$s \\
Bright Burst Rate & $\sim 1000$ per year (with measured $\ep$) \\
Flight Heritage & Fregate Instrument  on \hetetwo \\
Total Power &       25 W \\
Total Mass & 95 kg (50 kg Scintillators $+ 40$ kg PMTs $+5$ kg electronics) \\\hline
\end{tabular}
\end{center}
\label{table:scint_params}
\end{table*}

To refine these crude localizations and to extend the X-ray energy
range down to $\sim$1 keV, each facet also has eight large 
area Soft X-ray Cameras (SXCs) with X-ray CCD detectors
(Table \ref{table:sxc_params}).  The eight detectors
are equipped with different coded masks: 4 have one-dimensional masks
(2 oriented in the X direction and 2 in the Y direction), 2 have two-dimensional
masks, and 2 have no masks.  
The ensemble of eight SXC detectors works in concert
to provide excellent burst detection sensitivities in the 1-15 keV band,
as well as localization accuracy of 1.3\arcsec~in radius (90\% C.L.).  

\begin{table*}[htb]
\begin{center}
\caption{Silicon CCD Detectors for Prompt Satellite}
\begin{tabular}{ll}\hline\hline
Energy range &   1-15 keV \\
Field of View & 7 sr total (centered on anti-sun) \\
Module Configuration & 1D cameras: 2 X, 2 Y  \\
~(on 1 of 7 satellite faces)  & 2D cameras: 2-D,  2 Open \\
CCD Detectors & Deep depletion silicon; 72K $\times$ 72K, with 5 $\mu$m pixels \\
   & (Wafer scale device, segmented as $16 \times 16 = 256$ \\
  & independent ``panes'', each with $1152 \times 1152$ pixels) \\
Pixel Summation & 576 $\times$ 1 (1D) or $24 \times 24$ (2D) \\
Area per CCD &     90 cm$^2$ \\
Number of CCDs & 8 CCDs/Face $\times$ 7 Faces = 56 CCDs \\
Total CCD Area & 56 @ 90 cm$^2$ = 5040 cm$^2$ \\
CCD Temperature & Passive cooling to -100C \\
Coded Mask Size & 10 cm $\times$ 10 cm \\
Mask-to-Detector Separation &      20 cm \\
Camera  Resolution & 10.3\arcsec for Nyquist sampling of mask pattern \\
Source Localization Accuracy (90\% C.L.)  & 1.3\arcsec
radius for 5$\sigma$ detection in one XY
pair (2 CCDs) \\
Flux Sensitivity (10$\sigma$, 2-10 keV) &
              1 s: $4.5 \times 10^{-9}$ ergs cm$^{-2}$ s$^{-1}$ \\
            & 100 s:  $4.5 \times 10^{-10}$ ergs cm$^{-2}$ s$^{-1}$ \\
Time Resolution &     250 ms \\
Bright Burst Localization Rate & $\sim 800$ per year \\
Flight Heritage &  X-ray CCDs on {\it ASCA}, {\it Chandra},
{\it HETE-2}, \& {\it Astro-E} \\
Total Power &      100 W \\
Total Mass & 55 kg (40 kg cameras $+$ 15 kg electronics) \\\hline
\end{tabular}
\end{center}
\label{table:sxc_params}
\end{table*}

The
seven facets on the Prompt Satellite host 56 SXC cameras in total.
Each of the 56 SXC cameras contains a wafer-scale ($\sim$90 cm$^2$)
CCD consisting of 
16$\times$16 panes with 1152 $\times$1152 pixels each. The multipane approach has 
many advantages. Each charge packet moves only a small distance (1.2 
cm at most) to the readout, so that charge loss from 
radiation induced traps is minimized. Bad panes, caused by 
manufacturing defects or by micrometeorites that penetrate the camera 
entrance window, may be isolated and ignored: coded aperture imaging 
is insensitive to small gaps in focal plane coverage. Thus, a high 
manufacturing yield of acceptable wafers as well as a long detector 
life in orbit is assured. Each pane includes charge summation 
structures to make pixel summation fast and efficient. For the 1D 
modules,  576 rows are summed to produce a tall (2.9 mm) narrow (5 
$\mu$m) effective pixel shape. For the 2D and open modules, the sums 
will be 24x24, for a 120 $\mu$m square effective pixel. This reduces 
the readout load from 19 billion pixels to 33 million. A 200 Mpix/s 
readout system will allow 4 Hz readout of the entire array of modules 
with allowance for overhead. Thus, the CCD detector time resolution is 
improved to 250 ms, 5 times better than for the \hetetwo SXCs.

A large array of SXC modules
requires a high performance data acquisition and processing system. We
have previously investigated the design of such a system for a possible mission of
opportunity. It appears practical, with technology available in 2005, to
acquire and process X-ray CCD data for $\sim 1/60$ the power needed with
\hete
(1992) technology. \hetetwo only provides $\sim 10$W of power to support the
SXC, but the larger spacecraft  we envision should be able to provide at
least 100W to the SXC array,
so we anticipate a data acquisition and processing system
with at least 600 times the \hetetwo capacity supporting $\sim 200$ times the
silicon area. The extra capacity will provide better time resolution
and better real time imaging.
Additional details on the SXC camera and X-ray CCD detectors are given in
Table A.3.

\begin{table*}[htb]
\begin{center}
\caption{Resources for GRB Prompt Satellite}
\vspace{5mm}
\begin{tabular}{lll}\hline\hline
Item & Power (W) & Mass (kg) \\\hline
X-ray CCDs (5,040 cm$^2$) & 100 & 40 -- Cameras \\
 & -- & 15 -- Electronics \\
Scintillation Detectors & 25 & 50 -- NaI \\
~(12,250 cm$^2$) & -- & 40 -- PMTs \\
 & -- & ~5 -- Electronics \\
Attitude Control System & 15 & 10 \\
~(Momentum wheels, Star trackers, & & \\
~Thrusters, Electronics) & & \\
Command \& Data Handling & 5 & 5 \\
~(Electronics) & & \\
Power System (Batteries, Converters & 5 & 10 \\
~Solar panels) & & \\
X-Band System & 20 & 1 \\
VHF System & 4 & 1 \\
Propellant (Cold gas) & -- & 5 \\\hline
  Sums & 174 & 182 \\\hline
  Margin & 87 (50\%) & 182 (100\%) \\\hline
TOTAL (w/ Margin) & 261 W & 364 kg \\\hline
\end{tabular}
\end{center}
\label{table:sxcresources}
\end{table*}

\begin{figure}
\centerline{
\includegraphics[width=3.0in]{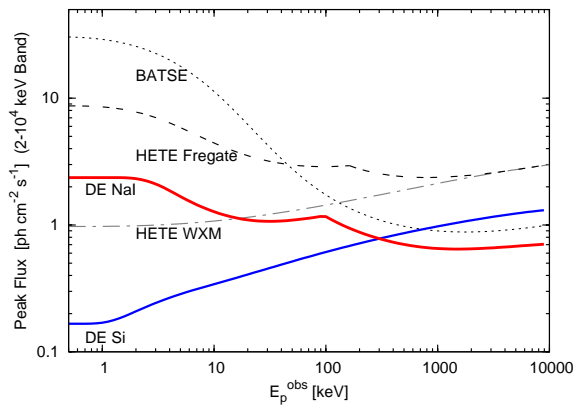}}
\caption{The sensitivity curves for the Dark Energy Si and NaI detectors
(colored lines) versus the observed GRB peak energy $\ep$.
Also plotted are the sensitivities for BATSE (dotted black line)
and \hetetwo (dashed black and dot-dashed gray lines).}
\label{fig:sens_grb}
\end{figure}

The sensitivity of the Prompt Satellite for establishing $\ep$ is depicted
in Figure \ref{fig:sens_grb}.  Both the scintillation detections (``DE NaI'') and
SXC cameras (``DE Si'') are more than an order of magnitude more sensitive
than flown on previous missions.

Despite the large areas of the NaI and SXC instrument arrays, they are compact,
low in mass, and require little power.  As a result, the spacecraft power
and mass resources are modest. We estimate that the Prompt Satellite
(a ``minisatellite'') will require a net of $<300$ W and $<400$ kg,
even allowing for ample reserves of 50\% and 100\%, respectively (Table A.3).

\subsection{X-ray Afterglow Satellite}

\begin{figure*}[htb]
\centerline{
\includegraphics[width=5in]{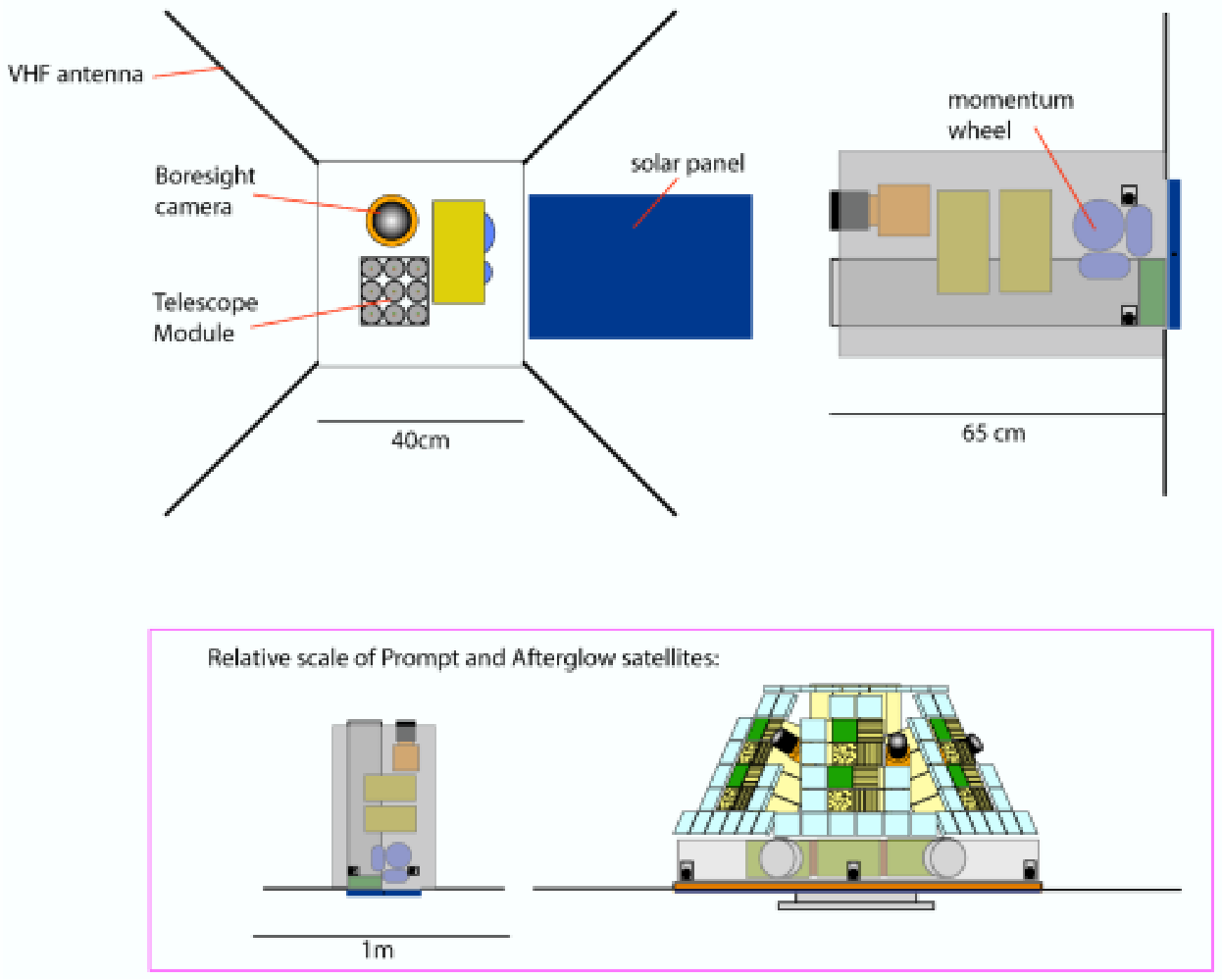}}
\caption{Schematic diagram of Afterglow Satellite (1 of 2) with
its X-ray Afterglow Telescope (XAT) array.}
\label{fig:schemdia1}
\end{figure*}

In Figure \ref{fig:schemdia1}, we show a concept for a microsatellite optimized
to observe the X-ray afterglows.  In Table A.4, we have
compiled the power and mass resources needed for each of two such microsatellites.
Even with a 100\% reserve, each Afterglow Satellite requires only 60\% of the
mass of {\it HETE-2}.  The power requirement per microsatellite is approximately
60\% of the power requirement of {\it HETE-2}.

\begin{table*}[htb]
\begin{center}
\caption{Resources for Afterglow Satellites (per Satellite)}
\vspace{5mm}
\begin{tabular}{lll}\hline\hline
Item & Power (W) & Mass (kg) \\\hline
CCD Detectors & 1 & 2 \\
Telescope Array & -- & 10 \\
Attitude Control System & 15 & 10 \\
~(Momentum wheels, Star & & \\
~trackers, Thrusters, Electronics) & & \\
 Command \& Data Handling & 5 & 5 \\
~(Electronics) & & \\
Power System & 4 & 8 \\
~(Batteries, Converters, & & \\
~Solar panels) & & \\
VHF System & 4 & 1 \\
Propellant (Cold gas) & -- & 2 \\\hline
Sums & 29 & 38 \\\hline
Margin & 14.5 (50\%) & 38 (100\%) \\\hline
TOTAL (w/ Margins) & 44 W & 76 kg \\\hline
\end{tabular}
\end{center}
\label{table:xrtresources}
\end{table*}

Each X-ray Afterglow Satellite will carry 9
($3\times3$) clustered, narrow field of view X-ray Afterglow Telescopes (XAT),
each with its own CCD detector.  The telescope and detector design parameters
and performance statistics for a typical observation (field angle of
0.5\arcmin) are displayed in Table \ref{table:xrt_params}.  The XAT design
is an updated version of the array of small telescopes flown and proven
aboard the SAS-3 satellite in the mid-1970s \citep{hearn76}. The
reflecting surface of each telescope is a paraboloid of rotation, coated with
Ni. A single bounce at grazing incidence focuses 0.5 keV X-rays onto the
CCD with excellent efficiency.  (The efficiency at 2 keV declines by only
a factor of 2 compared to its value near unity at 1 keV.)  Each telescope
is baffled to prevent X-ray's from field angles $>1\arcmin$ from reaching
the detector. The 9 telescope design is light ($<1$ kg per telescope), 
and the small
diameter (8 cm), thick-walled (3mm) unnested reflectors should be
inexpensive to fabricate.

\begin{table}[htb]
\begin{center}
\caption{X-ray Afterglow Telescope (XAT) Design Parameters}
\vspace{5mm}
\begin{tabular}{ll}\hline\hline
Focal Length ($F$)              &    80cm \\
Mirror Length                   &    60cm \\
Mirror Outer Diam.           &    8cm  ($F/10$) \\
Mirror Thickness                &    3mm \\
Number of Telescopes & \\
~per Satellite  &    9 \\
Mass per Telescope &    0.9 kg \\
Telescope PSF                   &    10.3\arcsec HEW, 0.5 keV \\
Effective Area                  &    286 cm$^2$, 0.5 keV \\
Detection Element               &    5x5 micron pixel \\
Pixel Scale                     &    1.3\arcsec/pixel \\
Field of View (radius)          & 60 arc second \\
Sensitivity (in 4 ksec)  & 2$\times$10$^{-14}$ erg/cm/s \\\hline
\tableline
\end{tabular}
\end{center}
\label{table:xrt_params}
\end{table}

\begin{figure}
\centerline{
\includegraphics[width=3.3in]{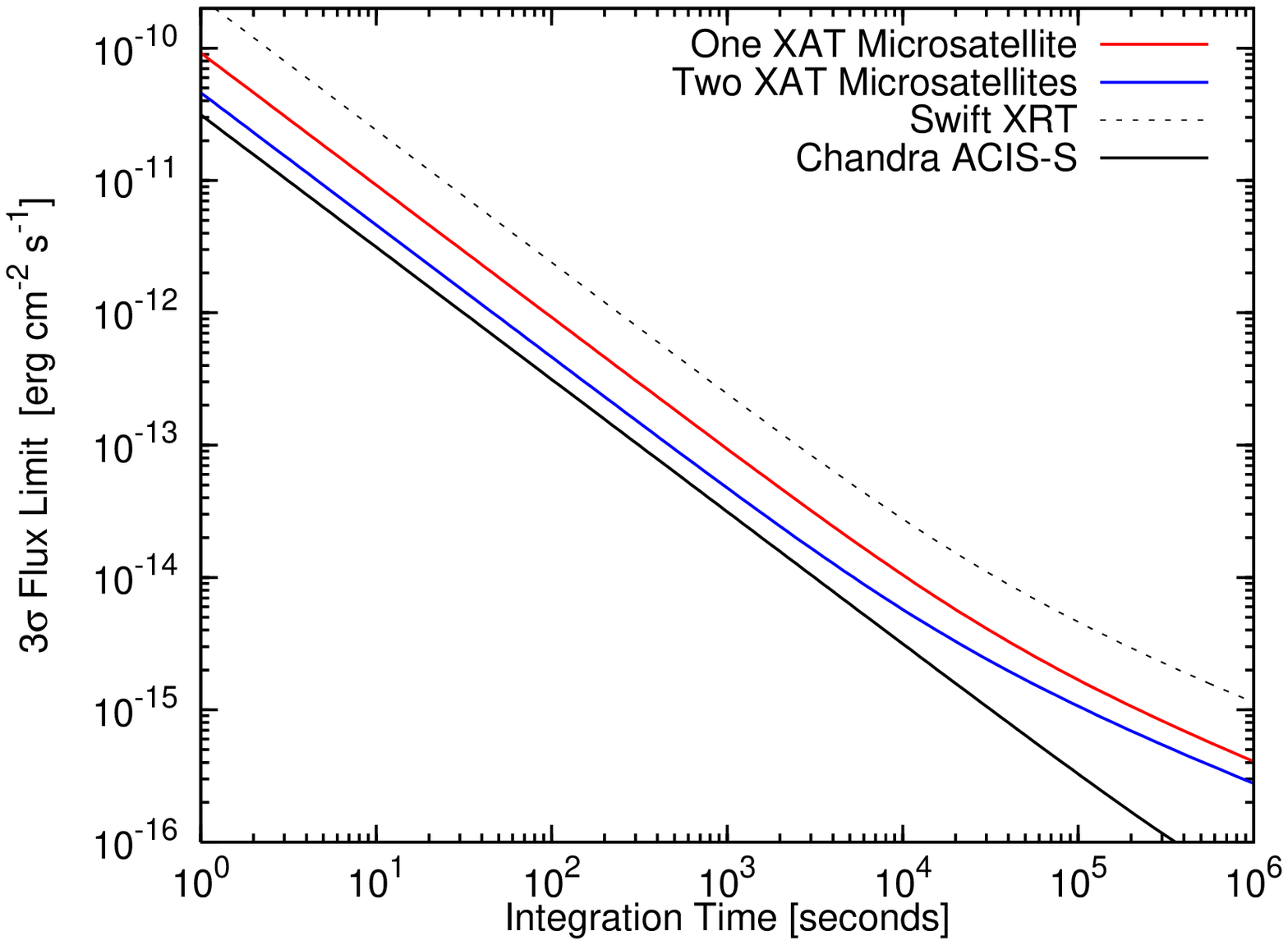}}
\caption{The XAT sensitivity (colored lines) versus exposure time.
The red line refers to a single XAT microsatellite viewing a given source,
while the blue line refers to both XAT microsatellites simultaneously
viewing the same source.  Also
plotted is the sensitivity for the \swift~XRT (dotted line) and
for {\it Chandra}~ACIS-S (heavy black line).}
\label{fig:xrt_sens}
\end{figure}

Each X-ray Afterglow Satellite will reach
a limiting sensitivity of $2 \times 10^{14}$ erg cm$^{-2}$ s$^{-1}$
in 4 ksec.  We take this flux limit as a conservative estimate for the
required sensitivity to detect late GRB jet break times. A 
flux $\sim 10\times$ greater was measured by the \swift~XRT for GRB~050408 at 
its jet
break ($t\sim 5$ days) after $\gtrsim 10$ ksec of integration.  Figure
\ref{fig:xrt_sens} shows the sensitivity of an XAT (solid line) relative
to the sensitivity of the \swift~XRT (dotted line).  For relatively
bright sources (fluence $\gtrsim 2 \times 10^{-14}$ erg cm$^{-2}$ s$^{-1}$), these
small XAT instruments achieve the same signal-to-noise determination as
does the \swift~XRT, yet the required exposure is $2.5\times$ shorter.
For fainter sources, or for sources in crowded fields, the required
exposure is $5-8\times$ shorter than for the \swift~XRT due to the
tighter PSF of the XAT (10.3\arcsec~HEW versus 
18\arcsec~PSF\footnote{http://www.swift.psu.edu/xrt/details}).

\subsection{Launch Vehicle Requirements}

\begin{figure*}
\centerline{
\includegraphics[width=5.94in]{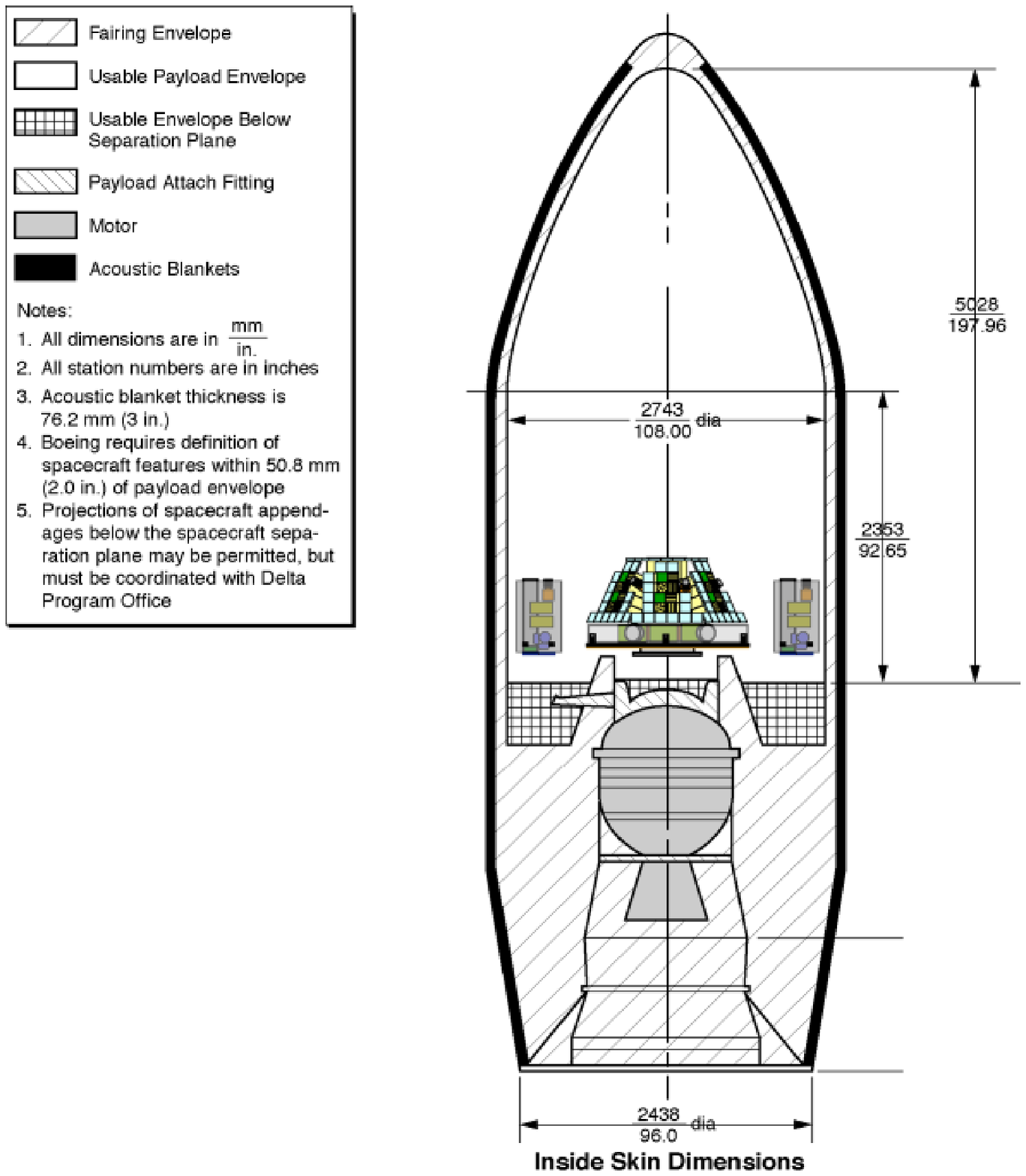}}
\caption{Schematic diagram of the GRB Prompt Satellite and 2 GRB Afterglow
Satellites in the Delta 7425 rocket faring.  There is ample clearance
for the 3 small satellites to fit within the faring.  Since the Delta 7425
can launch 800 kg to L$_2$, it can easily launch the combined payload mass
(see Tables A.3 and A.4), 
even with a mass reserve of more than 100\%.}
\label{fig:sat_fair}
\end{figure*}

As shown in Figure \ref{fig:sat_fair}, the Prompt Satellite and the two Afterglow
Satellites are easily accommodated within the fairing of a Delta 7425, the
same rocket used to launch the WMAP mission to L$_2$ \citep{bennet03}.  Based on
the mass estimates for the three satellites (Tables A.3 and A.4),
the Delta 7425 launch mass capability to L$_2$
of 800 kg provides ample ($>$100\% mass margin) for the MIDEX mission
we envision.

\subsection{Data Rate}

The data handling strategy for our mission combines concepts from \rxte and 
{\it HETE-2}. An {\it RXTE}-style bit code will represent raw event data from 
the scintillation detectors, with 32 channels of energy resolution and 20 ms 
time resolution. This code averages 2 bits/event. For the silicon detectors, 
the events are very sparse in the spatial/spectral/temporal resolution space. 
A run length encoding with 20 bits/event should be nearly optimal for these. 
The silicon detectors detect about an order of magnitude fewer photons than 
the scintillation detectors, so the total bit rates for the two detection 
systems are approximately equal.

Nearly all of the time, the diffuse X-ray background will dominate the 
event rate. A very intense burst may dominate the event rate for a few 
seconds, and a solar radiation storm may blind the detectors for awhile, 
but the diffuse X-ray rate determines the average raw data rate over time. 
For the encodings above, the average data rate is about 500 kbit/s. A 128 
Gbit data store will therefore be able to hold $\sim 3$ days of data on board to 
support ground queries for more detail.

Because 500 kbit/s is much larger than our raw downlink rate of 30 kbit/s, we 
will need to be selective in what we send to Earth. We will bin the raw data 
from each face of the localization/spectra spacecraft into 16 channels with 
1/4 second time resolution. With light data compression, this stream will be 
3.6 kbit/s. This will constitute a standard survey data set. For most 
bursts, this should be sufficient to determine fluence and spectral parameters.

The imaging software on board will conduct a spatially resolved search 
for bursts: we will devote 2 kbit/s to the results of this search.
The on board burst trigger algorithm will provide 1 kbit/s of diagnostic 
information.
Each of the two Afterglow Satellites will send 600 bit/s to the Prompt 
Satellite for relay to Earth.

The above, plus spacecraft housekeeping and data overhead, will use 10 
kbit/s, or 1/3 of the available downlink. The other 2/3 will be used to 
respond to ground queries for more detail. In most cases, we expect that all 
of the event data from a burst will be requested. For the brightest/longest 
bursts, an {\it RXTE}-like binning and recoding engine will reduce the 
data mass to a manageable amount.

For the afterglow satellites, 10 photon event/s will yield 320 bit/s. The 
remaining 280 bit/s will be housekeeping and overhead. In most cases the 
afterglow intensity will be much less than 1 photon/s at the jet break 
time so this should provide adequate dynamic range.

\begin{table}[htb]
\begin{center}
\caption{X-band Downlink Budget}
\vspace{5mm}
\begin{tabular}{ll}
\hline
Data type                       & kbit/s \\
\hline
Afterglow data relay           & 1.2 \\
GRB Survey                      &  3.6 \\
Soft imaging search     & 2.0 \\
Trigger status                  & 1.0 \\
Housekeeping                    & 1.0 \\
Overhead                        & 1.2 \\
Detailed analysis               & 20.0 \\
\hline
TOTAL                                   & 30.0 \\
\hline
\end{tabular}
\end{center}
\end{table}

\subsection{Navigation and Station-Keeping}

For simplicity, the Delta 7425 payload bus will transport the ``mother'' 
mini spacecraft (aka Prompt Satellite) and the accompanying
two \newline
``daughter''
micro spacecraft (aka Afterglow Satellites) to the vicinity of the
Earth-Sun L$_2$ Lagrangian point where they will be injected into a so-called
halo (or Lissajous)
orbit of {\it ca.} 100,000-200,000 km radius. Such orbits \citep{farquhar70} lie
close to a plane oriented at $\sim$$72^{\circ}$ from the ecliptic. Once
a halo orbit has been achieved, the daughter spacecraft will separate at
low relative velocities so as to execute Lissajous orbits \citep{barden98}
relative to the mother spacecraft. None of the spacecraft will be in
stable orbits---they will separate from each other, and from the vicinity
of the L$_2$ point, along an ``unstable manifold'' of the halo orbit at
an exponentially increasing rate unless appropriate velocity corrections
are made.
The mother spacecraft will contain an X-band transceiver, synchronous
transponder, and high-gain antenna, allowing it to be tracked directly
from Earth---utilizing round-trip time delay measurements to determine
range, and two-way 
Doppler tracking to measure line-of-sight velocity,
right ascension, and declination. Daily tracking as part of science data 
acquisition contacts at $\sim$100 m accuracy
will be sufficient to permit velocity corrections to be computed and
executed to maintain the halo orbit with low fuel consumption and slow
drift orthogonal to the orbit plane \citep{collange04}. Since the total
delta-V requirements will be  $< 5$ m s$^{-1}$ yr$^{-1}$ \citep{farquhar&dunham},
a small cold gas system ($<5$ kg propellant for the mother-sat, $<$2 kg 
propellant for each of
the daughter-sats) will suffice for  the nominal 2 year mission.
We propose the use of a simple low-gain phased-array antenna on the
daughter spacecraft to determine the direction to the mother spacecraft
with the VHF uplink; 4-6 whip antennas at different places on the
spacecraft body will feed a coherent multichannel digital receiver. The
phase difference between the antennas will indicate the direction of the
mother spacecraft. (With antennae separated by less than a wavelength this
is more like classic 1940's radio direction finding than a 21st century
phased array, but it should suffice for our requirements.) This knowledge
may also be used to phase the antennas for maximum satellite-to-satellite
link margin.
The daughter spacecraft will relay their data through the mother
spacecraft, and cannot therefore be reliably tracked from the
Earth. Instead, their range, range rate, and bearing from the mother
spacecraft will be determined using their VHF (2-meter, 140 MHz)
transceivers. Each daughter will possess a single dipole antenna, a VHF
transceiver, and a synchronous transponder, whereas the mother will employ
a 3 dimensional antenna array, {\it e.g.,} simple 1/4-wave VHF dipoles
at each apex of an equilateral pyramid, each with its own receiver. The
physical antenna can be spring-deployed (tape measure) elements of the
type used on the \hete satellite, for which the VHF antenna was literally
a $\sim 50$ cm long spring tape measure that uncoiled in orbit. The mother
spacecraft will transmit on a single dipole and its receiver outputs
will be heterodyned to baseband and cross-correllated to extract the
relative phase delay ($\pm 10^{-3}$ rad) of each baseline. Since the
antenna spacing is comparable to the carrier wavelength, aliasing will
not occur and the direction from the mother to each daughter will be
determined to $\sim$1 milliradian accuracy. Combined with traditional
VHF ranging and range-rate extraction, the relative location of each
spacecraft can be estimated, and velocity corrections can be computed
to maintain each daughter in the vicinity of the mother.

\subsection{Satellite Telecommunication}

The constellation of GRB Prompt (``mother'') and two Afterglow 
(``daughter'') Satellites will circle about L$_2$ in a halo orbit, of 
radius 100,000-200,000 km. All ground communication will pass through the 
Prompt Satellite over an X-band (10 GHz) link, utilizing two small 
($\sim 2000$ cm$^2$), flat phased array antennae on the sun- (and earth-) facing 
surface. The phased array antennae will permit continuous 
communication to earth without requiring a steerable dish or a 
re-orientation of the Prompt Satellite, which would disrupt GRB 
science data collection. The beam of the phased array antennae will 
be sufficiently narrow ($<2^{\circ}$) to avoid interference from solar 
RFI, yet broad enough to illuminate the entire earth (viewed from the L$_2$
halo orbit, the earth subtends $\sim 0.5^{\circ}$ and is offset $\sim 4^{\circ}$ 
from the sun). 
In addition, the Prompt Satellite will communicate with the 
Afterglow Satellites via 2-way VHF link.

\begin{table*}[htb]
\begin{center}
\caption{X-band Telemetry Link Margins for a 5m Diameter Ground-Based Antenna}
\vspace{5mm}
\begin{tabular}{cc}\hline
X-band Prompt Satellite-to-Ground & \\\hline\hline
Eb/N at 1e-6: & 7.6dB \\
Modulation:   &  BPSK$+$1/2Viterbi \\
Spectrum utilization: & 0.5 \\
Spacecraft TX power: & 5W \\
Distance: & 1,504,0000 km \\
Bit rate: & 30 kbps \\
Link margin: & 4.13 dB \\\hline
X-band Ground-to-Prompt Satellite & \\\hline\hline
Eb/No at 1e-6: & 14.2 \\
Modulation: & CPFSK \\
Spectrum utilization: & 1 \\
Ground station TX power: & 50W \\
Bit rate: & 113,600 bps \\
Link margin: & 5.20 dB \\\hline
\end{tabular}
\end{center}
\end{table*}

\begin{table}[htb]
\begin{center}
\caption{VHF Telemetry Link Margins}
\vspace{5mm}
\begin{tabular}{cc}\hline
VHF Satellite-to-Satellite & \\\hline\hline
Frequency: & 137.96 MHz \\
Spacecraft TX power: & 1 W \\
Bit rate: & 600 bps \\
Link margin  & 27 dB @ 10$^3$ km  \\
(@ satellite-to-satellite  &       7 dB @ 10$^4$ km  \\
~distance) & \\\hline
\end{tabular}
\end{center}
\end{table}

The VHF transceiver system described in ``Navigation and 
Station-Keeping'' will also serve to transmit the GRB burst alerts and 
$\sim$ arcsecond localizations from the Prompt Satellite to the Afterglow 
Satellites. (An automated scheduler on the Prompt Satellite will 
establish which of the two Afterglow satellites will be assigned to a 
given alert.) In addition, the VHF link will be utilized to transfer 
the X-ray afterglow observation data from the Afterglow Satellites 
back to the Prompt Satellite, for re-transmission to earth. We 
estimate that the satellite-to-satellite separations can be easily 
maintained to $<$1000 km, even if the station-keeping thruster events 
are only scheduled a couple of times per month. To assess the link 
requirements for this VHF system, we have scaled from the measured 
properties of the \hete VHF burst alert system. (NB: Although \hete is 
is low earth orbit and communicates to ground receivers, the link 
distances are typically 1000-2000 km, which is comparable to the 
satellite-to-satellite distances near L$_2$ that we are considering. 
Furthermore, the VHF RFI environment is much more favorable at L$_2$, 
since VHF RFI is dominantly terrestrial in origin. )

\setcounter{table}{0}
\renewcommand{\thetable}{B.\arabic{table}}
\setcounter{figure}{0}
\renewcommand{\thefigure}{B.\arabic{figure}}

\section{Instrument Concept for Ground-Based Segment -- Dedicated Telescope 
Network}
\subsection{Integral Field Spectrometer-equipped Telescopes}

A 2.4-meter Ritchey-Chretien telescope (such as production units from 
EOST\footnote{http://www.eostech.com/two\_point\_four\_meter.php})
will provide a
small 30 arcsecond diameter field of view to an lenslet array coupled to optical
fibers. An off-the-shelf 2.4m telescope should be able to slew and set 
to 3 arcsecond
accuracy. Focus and pointing will be monitored by pointing to a table
of known standard stars during clear night operations.

The lenslet
array will be a set of hexagonal elements, each subtending 0.5 square
arcseconds (approx 0.7\arcsec~on a side).  We require 800 lenslets to cover
a square 20\arcsec x20\arcsec spectroscopic FOV at Cassegrain focus. The output of
the fibers will feed a double-spectrograph, similar in design to the
SDSS grism spectrographs.  The fibers will be aligned and collimated to
provide an average of 5 pixel separation between individual spectra, and
2000 pixels on each blue and red side (4000 pixels total in each spectrum,
with minimal overlap near the dichroic).

\begin{table}[htb]
\begin{center}
\caption{Sensitivity Estimate for Spectrgraphic Redshift Limits for Ground-Based IFS-equipped Telescopes}
\vspace{5mm}
\begin{tabular}{cccc}\hline\hline
$m_{AB}$  & Bright Sky S/N &  Dark Sky S/N\\
   &   19 mag/arcsec$^2$ & 21 mag/arcsec$^2$ \\\hline
16.0 &  75.3 &  78.9 \\
17.0 &  43.4 &  48.2 \\
18.0 &  23.0 &  28.2 \\
19.0 &  11.0 &  15.4 \\
20.0 &   4.8 &   7.6 \\
21.0 &   2.0 &   3.4 \\\hline
\end{tabular}
\end{center}
\label{table:spectro}
\end{table}

If we assume pixels which are
approximately 100 km/s in dispersion, and an effective aperture of 1.0 m$^2$
(25\% end-to-end efficiency) then a single 900 second exposure will yield
approximately the S/N values given in Table \ref{table:spectro} for two extremes in sky conditions.

The spectrographs will have complete wavelength coverage from 3100-9600
Angstroms.  This covers the absorption line doublet of MgII over the
full redshift range of $0.11 < z < 2.4$. In addition, it will cover the
CIV doublet absorption from $1.0 < z < 5.0$, and Damped Lyman-$\alpha$
absorption from $1.6 < z < 6.5$.

\begin{table}[htb]
\begin{center}
\caption{Budget Breakdown for Dedicated Support Telescope \& IFS
Instrument Configuration}
\vspace{5mm}
\begin{tabular}{ll}\hline\hline
  \$2.0M &  2.4 meter automated telescope: \\
 & built,tested and delivered \\
  \$1.5M &  Optical double-sided spectrograph \\ 
 & with fiber/lenslet array. \\
  \$0.3M &  Estimate for site lease and minimal \\
 & observatory support over 2 years.  \\
  \$1.0M & Data reduction, $z$ determination \\
 & and quick response support \\
  & over the 2 year mission. \\\hline
  \$4.8M & TOTAL per site \\\hline
\end{tabular}
\label{table:costs}
\end{center}
\end{table}

To detect MgII and CIV robustly,
one would require an approximate S/N of 5 per pixel.  So in dark-sky
and bright-sky (3/4 moon) conditions, a single 900 second exposure is
sufficient down to $m_{AB}$ = 20th magnitude. To accurately detect these
MgII and CIV redshifts down to $m_{AB}$=21 will require 90 minutes
of exposure in bright time and 30 minutes in dark lunar 
conditions (see Table \ref{table:spectro}).
The Damped Lyman-alpha feature is much easier to detect, and should be
recovered in any spectra with S/N $>$ 2.0 per pixel.

To have complete sky coverage, we would
require 5 or 6 identical telescopes adequately separated
in longitude. Candidate sites include: Canary Islands, Chile, 
Southwestern USA, Hawaii, Australia, and Southern Africa. Each of the 
telescopes would be built at an established
astronomical site with good infrastructure.  Although the telescopes will
be designed to run autonomously, lost observing time could be minimized 
if observatory staff could
respond in a reasonable time if trouble arose. In Table \ref{table:costs}, 
we give the 
estimated costs per dedicated observatory site for a two year mission. 
Since the total is \$4.8M per site, establishing 6 sites would require 
$\sim$\$30M, which would comprise $\sim$10\% of the total budget for 
the MIDEX mission.


\clearpage
\begin{thebibliography}{}


\bibitem[Amati et al.(2002)]{amati2002}
	Amati, L., et al. 2002, A\&A, 390, 81

\bibitem[Amati et al.(2004)]{amati2004}
	Amati, L., et al. 2004, A\&A, 426, 415

\bibitem[Atteia(2005)]{atteia2005}
	Atteia, J.-L. 2005, private communication

\bibitem[Barbier et al.(2005)]{barbier2005}
	Barbier, L., et al. 2005, GCN Circular No. 3162	

\bibitem[Bloom et al.(2003)]{bloom2003} 
	Bloom, J.~S., Frail, D.~A., \& Kulkarni, S.~R.\ 2003, \apj,
	594, 674 

\bibitem[Firmani et al.(2005)]{fgga-v2005}
	Firmani, C., Ghisellini, G., Ghirlanda, G., \& Avila-Reese, V.\
	2005, MNRAS, 360, L1
	
\bibitem[Frail et al.(2001)]{frail2001} 
	Frail, D.~A., et al.\ 2001, \apjl, 562, L55

\bibitem[Freedman et al.(2001)]{freedman2001}
	 Freedman, W.~L., et al. 2001, \apj, 553, 47 

\bibitem[Friedman \& Bloom(2005)]{fb2005}
	Friedman, A. S. \& Bloom, J. S. 2005, ApJ, 627, 1

\bibitem[Fynbo et al.(2004)]{fynbo2004}
	Fynbo, J. P. U., et al. 2004, ApJ, 609, 962

\bibitem[Ghirlanda et al.(2004a)]{ggl2004} 
	Ghirlanda, G., Ghisellini, G., \& Lazzati, D.\ 2004a, \apj,
	616, 331 

\bibitem[Ghirlanda et al.(2004b)]{gglf2004} 
	Ghirlanda, G., Ghisellini, G., Lazzati, D., \& Firmani, C.\
	2004b, \apjl, 613, L13

\bibitem[Ghirlanda et al.(2005)]{gglf2005}
	Ghirlanda G., Ghisellini, G., Lazzati, D. \& Firmani, C.\ 2005,
	in Proceedings of the Fourth Rome Workshop on GRBs in the
	Afterglow Era, Il Nuovo Cimento, in press (astro-ph/0504184)

\bibitem[Hamuy et al.(1996)]{hamuy1996}
	Hamuy, M. et al. 1996, AJ, 112, 2398

\bibitem[Hu(2005)]{hu2005}
	Hu, W.\ 2005, ApJ, in press (astro-ph/0407158)

\bibitem[Kawai et al.(2005)]{kawai2005}
	Kawai, N., et al. 2005, GCN Circular No. 3013

\bibitem[Kelson et al.(2003)]{kelson2003}
	Kelson, D. D., et al. 2003, GCN Circular No. 2627

\bibitem[Lamb \& Reichart(2000)]{lr2000} 
	Lamb, D.~Q., \& Reichart, D.~E.\ 2000, \apj, 536, 1 

\bibitem[Lamb et al.(2004)]{lamb2004}
	Lamb, D.~Q., et al.\ 2004, New Astronomy, 48, 423

\bibitem[Liang \& Zhang(2005)]{liang2005}
	Liang, E., \& Zhang, B.\ 2005, ApJ, submitted (astro-ph/0504404)

\bibitem[Linder(2003)]{linder2003}
	Linder, E. V. 2003, PRL, 90, 091301

\bibitem[Perlutter(1999)]{perlmutter1999}
	Perlmutter, S., et al. 1999, ApJ, 517, 565

\bibitem[Riess et al.(2004)]{riess2004}
	Riess, A. G., et al.\ 2004, ApJ, 607, 665 (R04)

\bibitem[Rowan-Robinson(2001)]{rowan-robinson2001}
	Rowan-Robinson, M. 2001, APJ, 549, 745

\bibitem[Sakamoto et al.(2004)]{sakamoto2004}
	Sakamoto, T., et al. 2004, ApJ, 602, 875

\bibitem[Sakamoto et al.(2005a)]{sakamoto2005}
	Sakamoto, T., et al. 2005a, ApJ, in press (astro-ph/0409128)

\bibitem[Sakamoto et al.(2005b)]{sakamoto2005b}
	Sakamoto, T., et al. 2005b, ApJ, to be submitted

\bibitem[Sakamoto et al.(2005c)]{sakamoto2005c}
	Sakamoto, T., et al. 2005c, GCN Circular No. 3189

\bibitem[Soderberg et al.(2004)]{soderberg2004}
	Soderberg, A. M., et al. 2004, ApJ, 606, 994

\bibitem[Spergel et al.(2003)]{spergel2003}
	Spergel, D. N. et al. 2003, APJS, 148, 175

\bibitem[Wells et al.(2005)]{wells2005}
	Wells, A. A., et al. 2005, GCN Circular No. 3191

\end{thebibliography}

\begin{thebibliography}{}

\bibitem[Barden \& Howell(1998)]{barden98}
 Barden, B.~T., \& Howell, K.~C., 1998, AAS 98
\bibitem[Bennett et al.(2003)]{bennet03}
 Bennett, C.~L., et al. 2003, \apj, 583, 1
\bibitem[Berger et al.(2005)]{berger05}
 Berger, E., et al., 2005, astro-ph/0505107
\bibitem[Collange \& Leitner(2004)]{collange04}
 Collange G., \& Leitner, J., 2004, AIAA 2004-4781, Providence, 16-19 August
\bibitem[Doty(2004)]{doty04}
 Doty, J. 2004, AIP Conf. Proc. 727, 708 
\bibitem[Farquhar(1970)]{farquhar70}
 Farquhar, R.~W. 1970, NASA technical report, R-346
\bibitem[Farquhar \& Dunham(1990)]{farquhar&dunham}
 Farquhar, R.~W., \& Dunham, D.~W. 1990, 
 Observatories in Earth Orbit and Beyond, Y. Kondo,
 ed., Klewer Academic Publishers, 1990, p 391
\bibitem[Hearn et al.(1976)]{hearn76}
 Hearn, D.~R., et al. 1976, \apj, 203, L21
\end{thebibliography}
\end{document}